%% file: main.tex
\documentclass[letterpaper,twocolumn,10pt]{article}
\usepackage{usenix}

\usepackage{amsmath}

\usepackage{filecontents}

\usepackage{authblk}
\usepackage{fontawesome}
\usepackage{pifont}
\usepackage{subfig}
\usepackage{appendix}
\usepackage{multirow}
\usepackage{varwidth}
\usepackage{tabularx}
\usepackage{graphicx}
\usepackage{makecell}
\usepackage{enumitem}
\usepackage{parskip}
\usepackage{circledsteps}
\usepackage[skins]{tcolorbox}

\pgfkeys{/csteps/inner color=white}
\pgfkeys{/csteps/fill color=black}
\pgfkeys{/csteps/outer color=black}

\newcommand{\etal}{\textit{et al}. }

\newcommand{\VoltSchemer}{\emph{VoltSchemer}}
\newcommand{\cmark}{\ding{51}}%
\newcommand{\xmark}{\ding{55}}%
\newcommand{\hliwona}[1]{{\fontfamily{iwona}\selectfont \textbf{#1}}}

\definecolor{5W}{HTML}{F1FAEE}
\definecolor{10W}{HTML}{669BBC}
\definecolor{15W}{HTML}{CA3402}
\definecolor{SCBox}{HTML}{F4F1DE}
\definecolor{Insight}{HTML}{81B29A}
\definecolor{InsightLight}{HTML}{C0D8CC}

\makeatletter
\newcommand{\DrawLine}{%
  \begin{tikzpicture}
  \path[use as bounding box] (0,0) -- (\linewidth,0);
  \draw[color=red!75!black,line width=0.5mm]
        (0-\kvtcb@leftlower-\kvtcb@boxsep,0)--
        (\linewidth+\kvtcb@rightlower+\kvtcb@boxsep,0);
  \end{tikzpicture}%
  }
\makeatother

\begin{document}

\date{}

\title{ \VoltSchemer: Use Voltage Noise to Manipulate Your Wireless Charger}

\author[1]{Zihao Zhan}
\author[1]{Yirui Yang}
\author[1,2]{Haoqi Shan}
\author[1]{Hanqiu Wang}
\author[1]{Yier Jin}
\author[1]{Shuo Wang}
\affil[1]{University of Florida}
\affil[2]{CertiK}
\affil[ ]{\textit {\{zhan.zihao, yirui.yang, haoqi.shan, wanghanqiu\}@ufl.edu}}
\affil[ ]{\textit {yier.jin@ieee.org}}
\affil[ ]{\textit {shuo.wang@ece.ufl.edu}}

\maketitle
\input{files/abstract.tex}

\input{files/introduction}

\input{files/background}

\input{files/threat_model}
\input{files/attack_principles}

\input{files/implementation}
\input{files/VoiceInjection}

\input{files/charging_manipulation}
\input{files/foreign_object_destruction}

\input{files/discussion}

\input{files/related_work}

\input{files/ethical_considerations}

\input{files/conclusion}
\input{files/acknowledgement}

\bibliographystyle{plain}
\bibliography{files/reference.bib}

\appendices
\input{files/appendix-burned_marks}

\end{document}

%% file: files/abstract.tex
\begin{abstract}

Wireless charging is becoming an increasingly popular charging solution in
portable electronic products for a more convenient and safer charging experience
than conventional wired charging. However, our research identified new
vulnerabilities in wireless charging systems, making them susceptible to
intentional electromagnetic interference. These vulnerabilities facilitate a set
of novel attack vectors, enabling adversaries to manipulate the charger and
perform a series of attacks.

In this paper, we propose \VoltSchemer, a set of innovative attacks that grant
attackers control over commercial-off-the-shelf wireless chargers merely by
modulating the voltage from the power supply. These attacks represent the first
of its kind, exploiting voltage noises from the power supply to manipulate
wireless chargers without necessitating any malicious modifications to the
chargers themselves. The significant threats imposed by \VoltSchemer\ are
substantiated by three practical attacks, where a charger can be manipulated to:
control voice assistants via inaudible voice commands, damage devices being
charged through overcharging or overheating, and bypass Qi-standard specified
foreign-object-detection mechanism to damage valuable items exposed to intense
magnetic fields.

We demonstrate the effectiveness and practicality of the \VoltSchemer\ attacks
with successful attacks on 9 top-selling COTS wireless chargers. Furthermore, we
discuss the security implications of our findings and suggest possible
countermeasures to mitigate potential threats.

\end{abstract}

%% file: files/introduction.tex
\section{Introduction}

Given the widespread use of mobile devices that require daily charging, ensuring their charging security has become critical.
Numerous attacks have been explored to target the charging process through cables, allowing attackers to control devices, install malware, induce touch events, inject voice commands, and compromise user privacy~\cite{lau2013mactans, nohl2014badusb, Wang:2022:NDSS, Jiang:2022:SP, shiroma2017threat}. 
Most attacks affect primarily wired charging systems because they exploit the vulnerability of data wires in USB charging cables to conduct unauthorized data transmission with malicious power sources.
Wireless charging, however, not only offers more convenient charging experiences but also inherently resists many attacks commonly existing in wired charging systems.

Wireless charging uses near-field magnetic coupling for power transfer, eliminating the need for direct electrical connections to the charged device. 
This feature prevents malicious attackers from accessing the direct data pathway to the charged device, even if the power supply is compromised. 
Moreover, wireless power transfer processes are secured by enforcing adherence to the Qi standards developed by the Wireless Power Consortium (WPC)~\cite{Van:2010:PEMC}.
Qi standards incorporate robust safety mechanisms to protect both the charged device and other objects from potential damages imposed by the intense magnetic fields. 
The benefits of wireless charging, including enhanced security, simplified charging, extended device longevity, and reduced clutter, have led to its widespread adoption and ease of deployment.
Consequently, in recent years, the wireless charging market has rapidly expanded at a compound annual growth rate (CAGR) of 25.8\%~\cite{fortune2023wireless}.
Wireless chargers are now widely deployed in various public places such as airports, restaurants, hotels, and coffee shops.

However, despite their numerous benefits, our research identifies new, critical vulnerabilities that can be exploited to invalidate the security characteristics of wireless charging systems and launch powerful attacks.
Specifically, the schemed voltage noises from the power adapter can propagate through the power cable and modulate the power signals on the charger's transmitter coil due to the effects of electromagnetic interference (EMI) on the charger. 
This process directly modifies the power signal used for power transfer, opening the door for potential breaches. 
Qi wireless charging relies on in-band communication, in which the charger and the device exchange essential Qi messages through the direct modulation of the power signal. 
Therefore, an attacker can potentially control this communication channel by injecting finely-tuned voltage noises, thereby gaining the ability to instruct the charger to execute various malicious tasks.

In this paper, we introduce \VoltSchemer\ attacks that exploit the newly identified vulnerabilities. These attacks enable an attacker to \textbf{gain complete control over wireless chargers using intentional electromagnetic interference (IEMI) via the voltage supplied by a connected power source}. 
\VoltSchemer\ can modulate the strong magnetic field generated by the charger based on power electronics and EMI principles. 
This manipulation enables attackers to control smartphones' voice assistants by inducing unintended voice commands in their microphone circuits through near-field magnetic coupling.
In addition, \VoltSchemer\ can deceive a connected wireless charger with fabricated Qi messages, instructing it to initiate hazardous power transfers. 
These harmful power transfers can potentially damage
the charged device or other valuable items susceptible to intense magnetic
fields. To further validate the effectiveness of the \VoltSchemer\ attacks, we
conducted an evaluation on 9 top-selling Commercial-Off-The-Shelf (COTS)
wireless chargers. The results show that all the tested chargers are vulnerable
to our \VoltSchemer\ attacks, highlighting their broad risks and potential
impacts.

To summarize, the main contributions of this paper are:
\begin{itemize}[topsep=0pt,itemsep=0ex,partopsep=1ex,parsep=1ex]
    \item
          Through a comprehensive examination of the Qi wireless charging
          design, we discovered new vulnerabilities in its design and protocol.
          These vulnerabilities allow an attacker to gain full control over the
          charger by merely manipulating the power supply.

    \item
          We developed \VoltSchemer, a suite of novel attacks that capitalize on
          these newly identified vulnerabilities, utilizing an interposed
          voltage manipulator to interfere with the power adapter's output voltage.
          This allows potential
          attackers to commandeer the connected wireless chargers and engage in various harmful activities.

    \item
          We illustrated the potential threats of \VoltSchemer\ via three
          attacks: voice assistant manipulation, wireless power toasting, and
          foreign object destruction.\footnote{Readers can view our practical
          attack scenarios and associated video clips by visiting
          \url{https://sites.google.com/view/voltschemer/}} 

    \item
          We conducted extensive experiments for \VoltSchemer\ attacks on
          popular COTS wireless chargers. Our findings showcase the real-world
          applicability and the significant threats that our attacks pose.

    \item
          We discussed the security implications of our findings and proposed
          countermeasures to mitigate these threats.
\end{itemize}

%% file: files/background.tex
\section{Background}
\label{sec:background}

\subsection{Qi Wireless Charging}\label{sec:qi_wireless_charging}

\begin{figure}[htbp]
    \centering
    \includegraphics[width=\columnwidth]{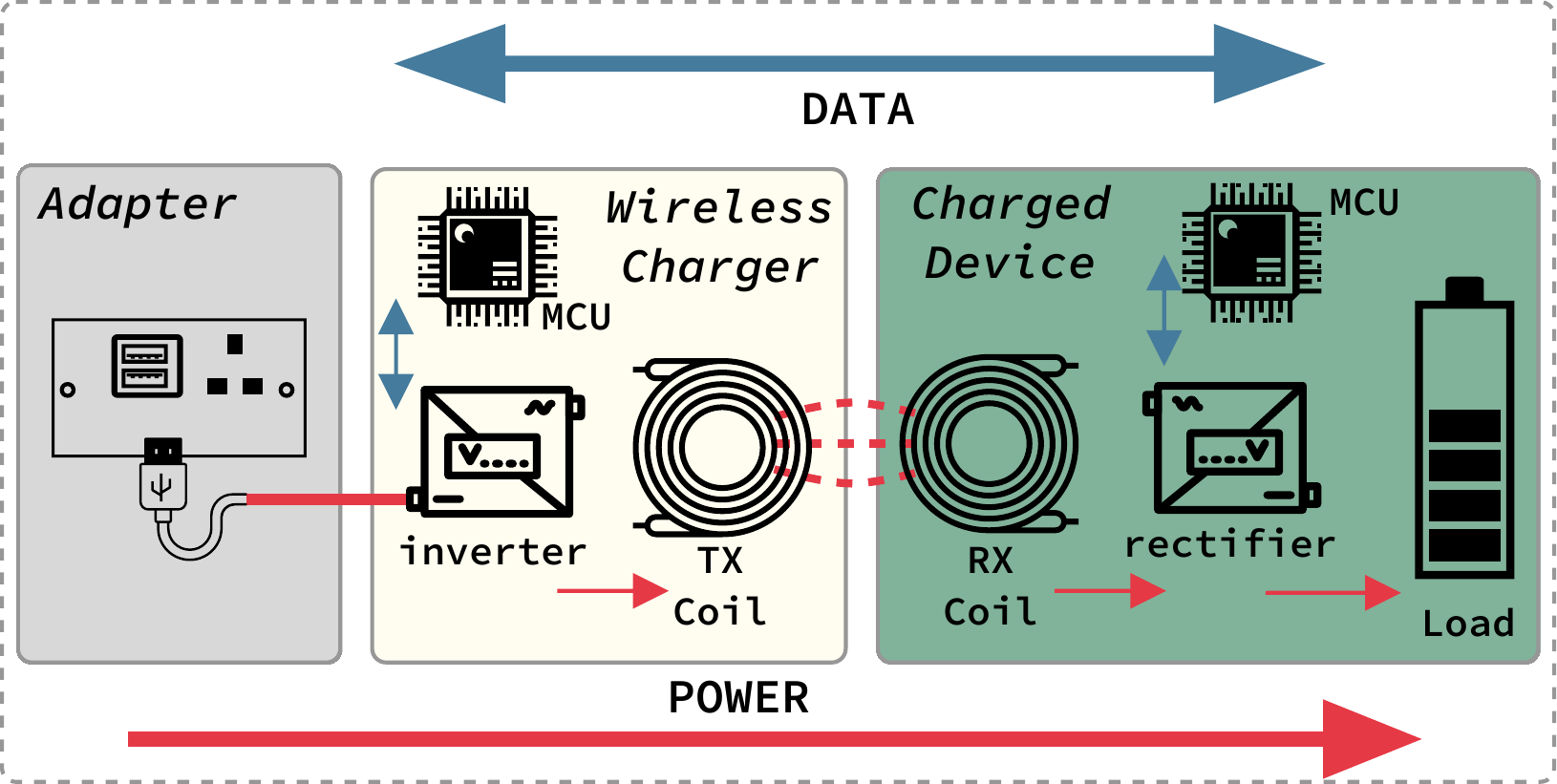}
    \caption{Overview of Wireless Charging System}
    \label{fig:background}
\end{figure}

A Qi wireless charging system comprises three primary devices depicted in
Figure~\ref{fig:background}: a power adapter, a wireless charger, and a charging
device. The power adapter's main function is to supply DC voltage to the
wireless charger via a power cable, such as a USB cable. The wireless charger,
also known as the power transmitter (TX device), utilizes an inverter to
convert the DC voltage from the power adapter into AC voltage on the TX coil.
The microcontroller unit (MCU) in the charger controls the amplitude and
frequency of this AC voltage, generating a strong alternating magnetic field
known as the power signal in wireless charging systems. The charged device, or
power receiver (RX device), captures this magnetic field through the RX coil,
inducing an AC voltage. The RX device's rectifier then converts this AC voltage
back into DC voltage and provides power to load.

\begin{figure*}[htbp]
    \centering
    \includegraphics[width=\textwidth]{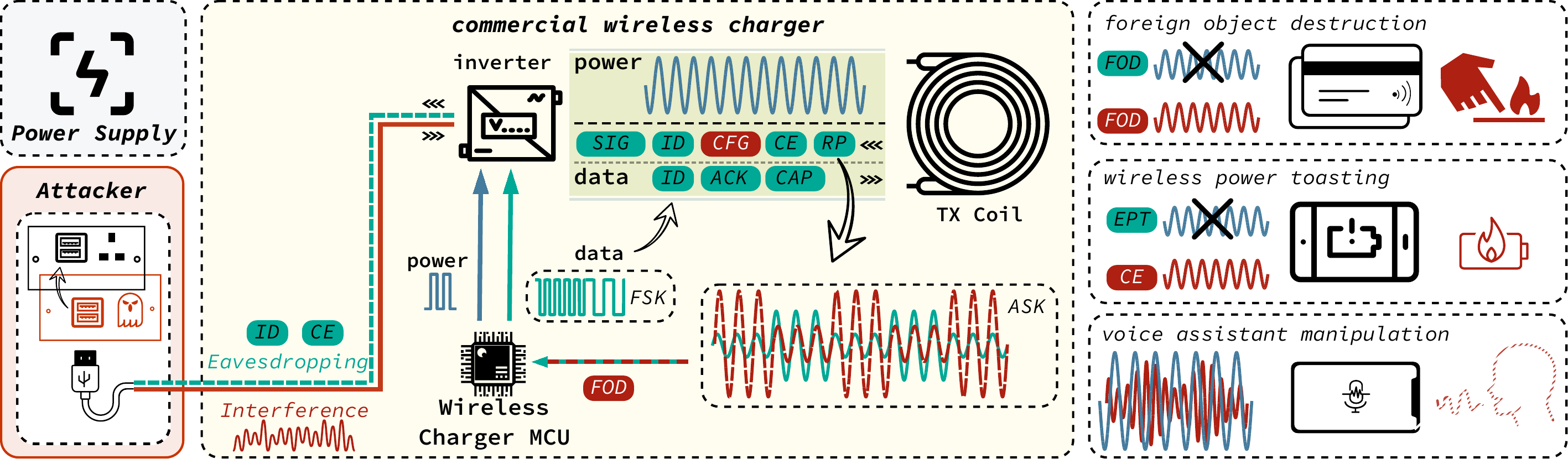}
    \caption{ 
    Attack overview: 
    A victim uses Commercial-Off-The-Shelf Qi-compatible wireless chargers and power receivers. 
    An intermediary-connected attacking device on the power adapter manipulates the output voltage and current to: 
    1) manipulate the magnetic field to interfere with the charged device. 
    2) interactively communicate with the charger and control the charging process. 
    This setup enables foreign object destruction, wireless power toasting, and voice assistant manipulation attacks. 
    }
    \label{fig:threat_model}
    \vspace*{-.15in}
\end{figure*}

One of the most significant distinctions between wireless and wired charging is
the absence of physical electrical connections to the RX device during charging.
A common vulnerability in wired charging is that electrical connections to a
charged device can inadvertently allow malicious actors to gain unauthorized
access to the charged device through the data wires in the charging
cable~\cite{lau2013mactans, Wang:2022:NDSS, shiroma2017threat}. Wireless
charging effectively eliminates this direct data path introduced by physical
connections. Therefore, an important \textbf{Security Characteristic (SC)}
provided by wireless charging is:\\
\begin{tcolorbox}[
    skin=widget,
    left=0mm,
    right=0mm,
    top=0mm,
    bottom=0mm,
    boxrule=1mm,
    colframe=SCBox,
    colback=SCBox,
    width=(\linewidth),
    nobeforeafter,
    ]
    \hliwona{SC 1: It eliminates physical connections to a charged device, thereby reducing its attack surfaces.}  
\end{tcolorbox}
Qi wireless charging also features robust in-band communication, where RX and TX devices exchange data by modulating and demodulating power signals using different schemes.
RX devices modulate power signals with Amplitude-Shift Keying (ASK) from the load side, while TX devices apply Frequency-Shift Keying (FSK) to modulate signals from the charger side.
Numerous techniques are specified to ensure communication robustness. 
For instance, Qi wireless charging uses Biphase Mark Coding (BMC) for bit encoding, which is known for its resilience to interference. 
Additionally, error detection bits and checksum bytes are incorporated to ensure data integrity. 
The robust Qi communication is crucial for the Qi standards' key
safety features, such as feedback charging control and foreign object detection,
ensuring a safe charging process.

\noindent\textbf{Feedback Charging Control} During charging, a power receiver
regularly sends Control Error (\texttt{CE}) packets to command the transmitter
to adjust the charging power. In response, the transmitter feeds the \texttt{CE}
value to a PID controller to update the controlling signal on the inverter. This
feedback control is essential to guarantee the charging power is
dynamically adjusted to meet the power receiver's needs. Furthermore, when the
power receiver detects abnormal charging status or is fully charged, it sends the End Power Transfer (\texttt{EPT}) packet to command the transmitter to terminate the charging.
Therefore, the second security characteristic provided by wireless charging
is:\\
\begin{tcolorbox}[
    skin=widget,
    left=0mm,
    right=0mm,
    top=0mm,
    bottom=0mm,
    boxrule=1mm,
    colframe=SCBox,
    colback=SCBox,
    width=(\linewidth),
    nobeforeafter,
    ]
    \hliwona{SC 2: It incorporates Qi communication-based feedback control to establish a safe charging process, thereby improving the longevity of charged devices.}  
\end{tcolorbox}
\noindent\textbf{Foreign Object Detection} 
Qi standards define Foreign Object Detection (\texttt{FOD}) to avoid heating and damaging magnetic-field sensitive foreign objects exposed in the magnetic field, enhancing the charging safety.
The \texttt{FOD} can be performed before and during the
power transfer. Pre-power transfer is mandatory when the power receiver requests
a high charging power using the extended power protocol. 
During this process, the power receiver sends a \texttt{FOD} packet containing the reference value of resonance properties to the transmitter. 
The transmitter compares this reference value with the value measured by itself to determine whether a foreign object is present. 
In-power transfer \texttt{FOD} is employed in both baseline and extended power protocols.
During charging, the power receiver must update the transmitter with the
Received Power (\texttt{RP}) packets. The power transmitter compares the
transmitted power measured by itself with the reported power received by the power receiver to
calculate the amount of unintended power transfer to foreign objects. If the
difference exceeds a predefined threshold, the charger identifies it as unsafe
and terminates the power transfer. Therefore, another security characteristic of
wireless charging is:\\
\begin{tcolorbox}[
    skin=widget,
    left=0mm,
    right=0mm,
    top=0mm,
    bottom=0mm,
    boxrule=1mm,
    colframe=SCBox,
    colback=SCBox,
    width=(\linewidth),
    nobeforeafter,
    ]
    \hliwona{SC 3: It specifies the FOD mechanism to restrict power transfer to foreign objects, thereby enhancing the safety and usability of wireless charging. }  
\end{tcolorbox}

%% file: files/threat_model.tex
\section{Threat Model and Attack Overviews}\label{sec:threat_model}
Our threat model and attack scenarios are depicted in Figure~\ref{fig:threat_model}. 
We assume a commonly adopted threat model for charging attacks, where an
adversary compromises the power adapter that supplies DC voltages to the
wireless charging system. 
To achieve this, an attacker connects a disguised voltage manipulation device between the power adapter and wireless charger, inducing voltage fluctuations to manipulate the power signal via the EMI effect, enabling a series of attacks.
We do not presuppose the necessity for attackers to interfere with data
transmission lines in USB cables. The attacks are initiated when a victim
unsuspectingly leaves a smartphone or metallic personal items near the charging
area either for charging or non-charging purposes. The attacks listed below can invalidate all three security characteristics
introduced in Section~\ref{sec:background}.
\begin{tcolorbox}[
    skin=widget,
    left=0mm,
    right=0mm,
    top=0mm,
    bottom=0mm,
    boxrule=0.75mm,
    colback=red!5,
    colframe=red!75!black!80,
    width=(\linewidth),
    segmentation style={solid, line width=0.35mm, left color=blue!15!yellow, right color=red!85!black, dashed},
    title=\hliwona{Attack Overviews},
    ]

    \hliwona{Attack 1: An attacker can modulate the high-power magnetic field
    to inject voice commands into charged smartphones and manipulate the voice
    assistants.}  
    
    \tcbline
    
    \hliwona{Attack 2: An attacker can intercept the communication between RX
    and TX devices to induce a hazardous charging process that impairs the
    charged device.}  
    
    \tcbline
    
    \hliwona{Attack 3: An attacker can initiate unsafe power transfer to
    metallic foreign objects in close proximity to cause irreversible damage.}  
    
\end{tcolorbox}

%% file: files/attack_principles.tex
\section{Wireless Charging System Security
Analysis}\label{sec:vulnerability_analysis} To understand why and how attacks
can be carried out through the power cable of a wireless charging system, two
critical questions must be answered: 
\Circled{\textbf{1}} 
\textbf{How can interference impact a wireless charging system through its power cable, and in what ways?}
\Circled{\textbf{2}} \textbf{What detailed information regarding the status of a wireless
charging system can be collected from the power cable?} 

\begin{figure}[htbp]
    \centering
    \includegraphics[width=\columnwidth]{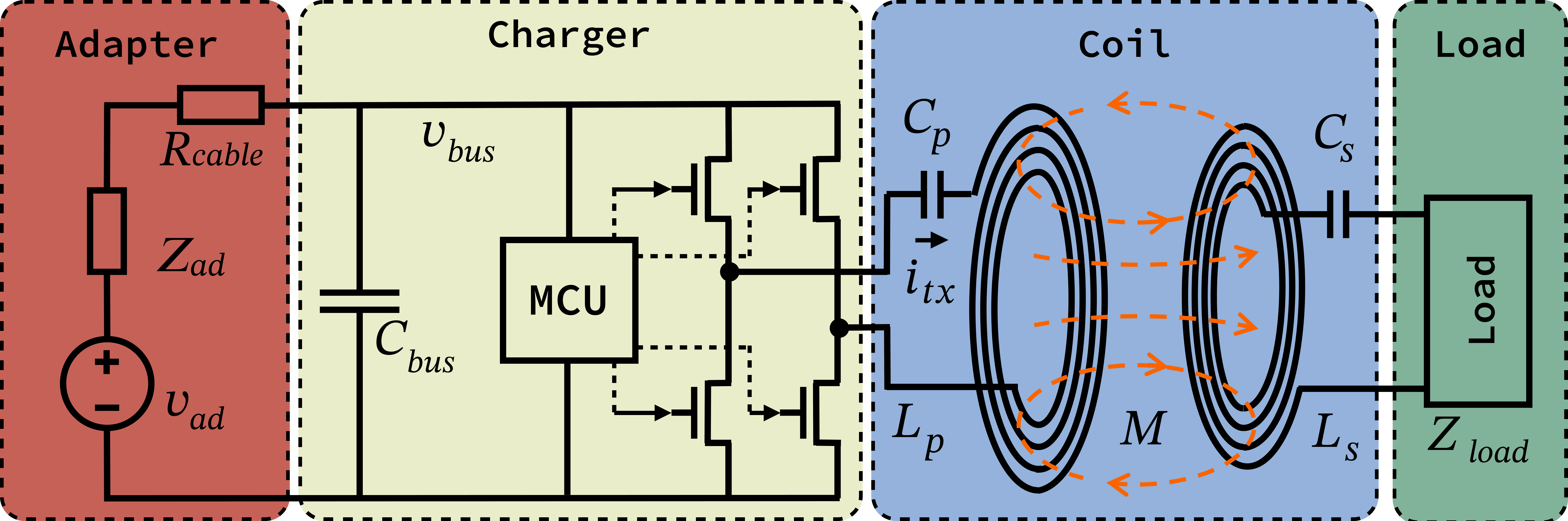}
    \caption{The schematic of a wireless charging system}\label{fig:WPT_Schematic}
\end{figure}

To answer these questions, we conducted a comprehensive analysis of the wireless charging system
depicted in Figure~\ref{fig:WPT_Schematic}. In Section~\ref{sec:fwdprop}, we examine how the schemed voltage interference at the power adapter's output propagates in the systems and impacts the transmitted power signal of the system. In Section~\ref{sec:backprop}, we explore how the workload
behavior-induced signals propagate back to the power adapter's output and impact the output voltage.

\subsection{
Adapter-to-Load Propagation
}\label{sec:fwdprop}

\begin{figure}[htbp]
    \centering
    \includegraphics[width=0.65\columnwidth]{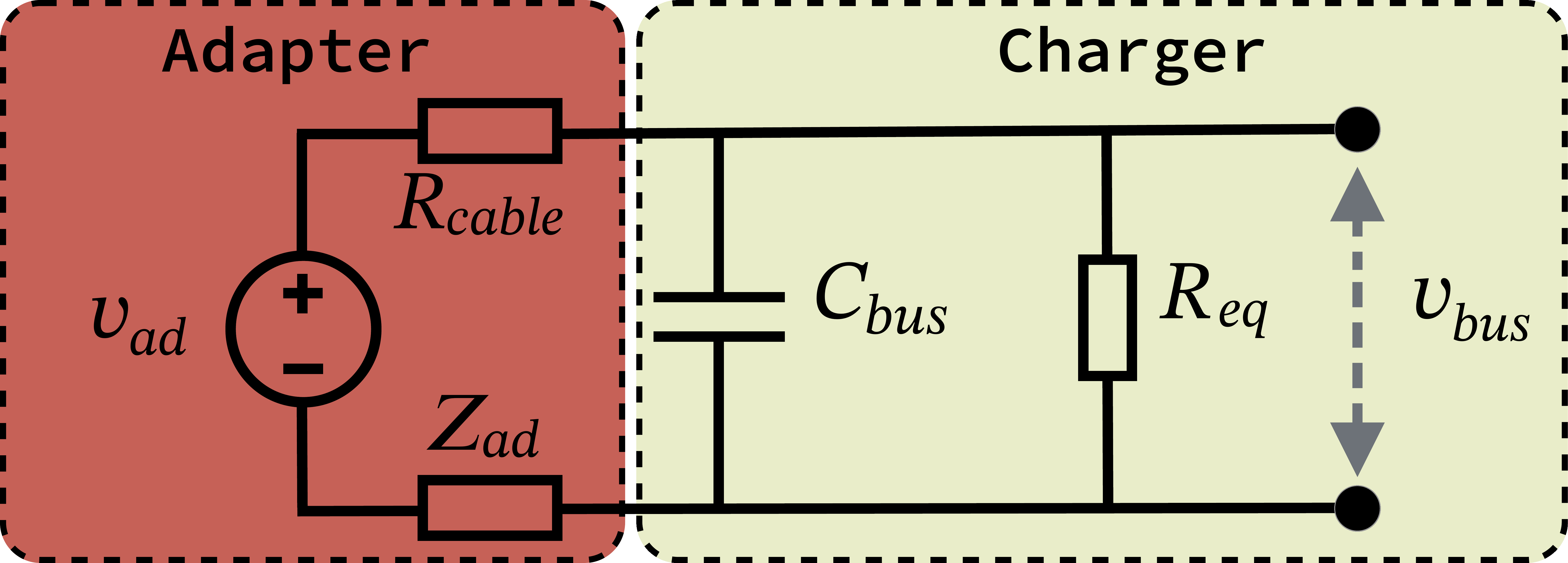}
    \caption{Circuit model to analyze the impact of power adapter's output voltage $v_{ad}$ on bus voltage $v_{bus}$}\label{fig:WPT_part1}
\end{figure}

\begin{figure}[htbp]
    \centering
    \includegraphics[width=0.65\columnwidth]{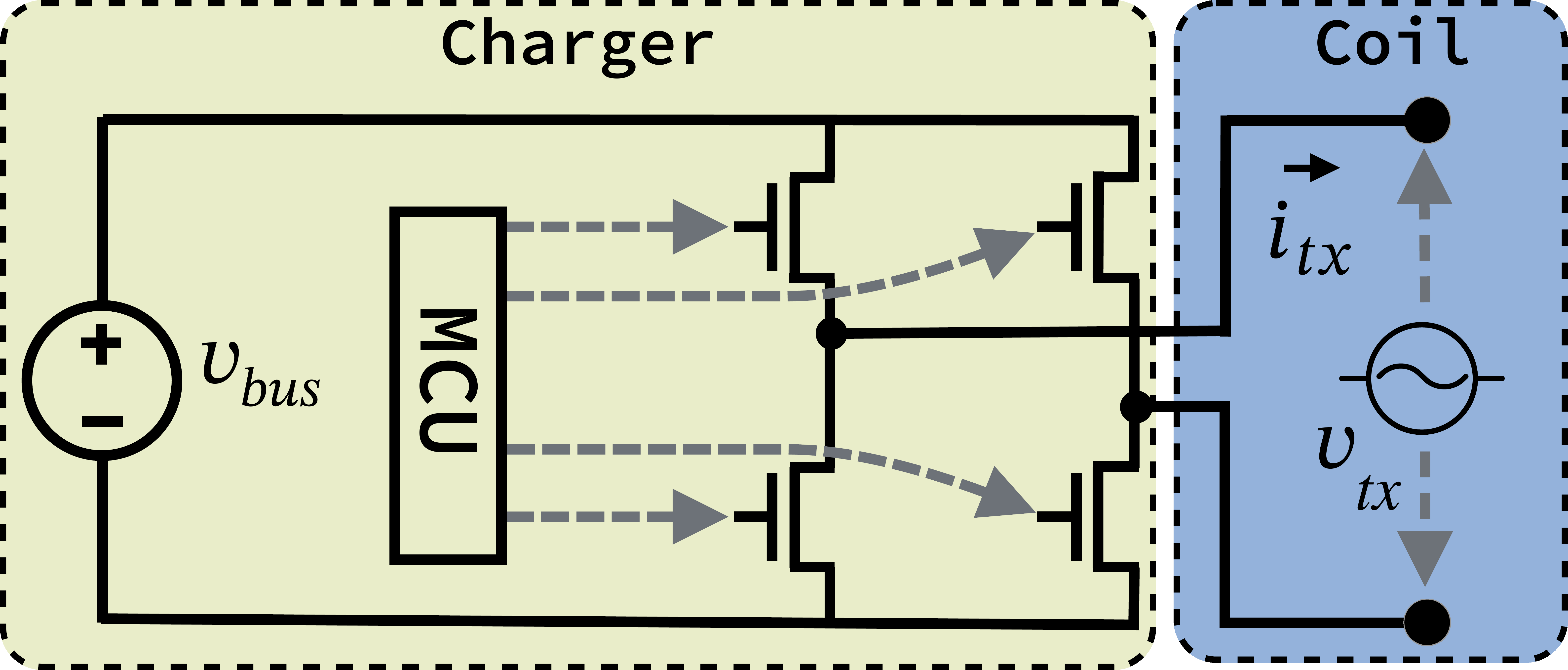}
    \caption{DC-AC inverter schematic}\label{fig:WPT_part2}
\end{figure}

\begin{figure}[htbp]
    \centering
    \includegraphics[width=0.65\columnwidth]{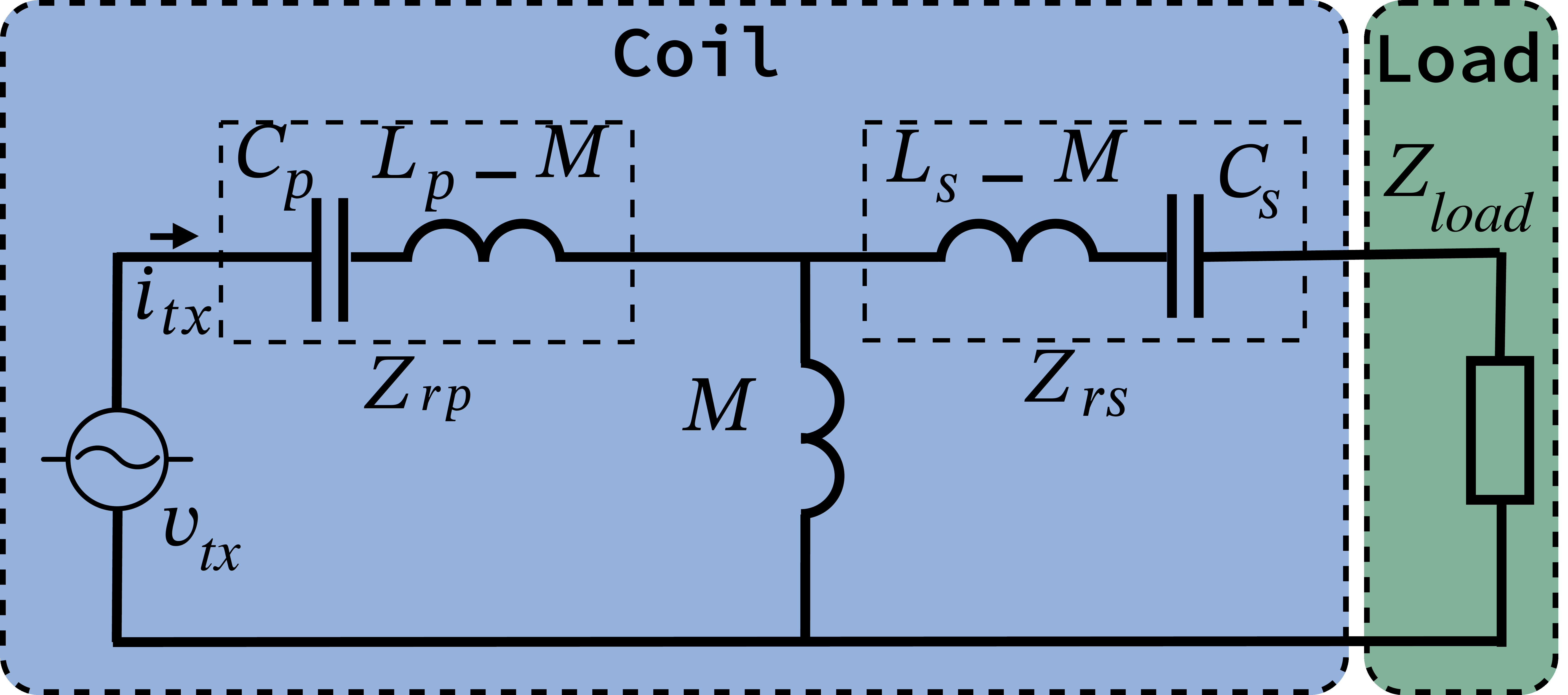}
    \caption{Circuit model for wireless power transfer analysis}\label{fig:WPT_part3}
\end{figure}

A regular wireless charging system follows electromagnetic compatibility and power electronics principles: ensuring that the noise from a power supply, a power adapter in this case,  does not disrupt the system's normal power conversion. However, the in-band communications employed in Qi wireless charging systems may encounter a different story. 
This section analyzes how an interference signal at the output of a power adapter affects the in-band communication, which is realized by modulating power signal transferred to the charging receiver via the couplings between the coils. 
We consider a scenario where the output voltage $v_{ad}$, as defined in Equation~\ref{eq:setup}, of an interfered power adapter is composed of the nominal DC output voltage $V_{ad}$ superimposed by a noise with an interference depth $m_i$ and frequency $f_i$,
\begin{equation}\label{eq:setup}
    \begin{split}
        v_{ad}(t) = V_{ad}(1  + m_i sin(2\pi f_i t)),\\
    \end{split}
\end{equation}
Because of large number of electronic components, including multiple non-linear components such as time-variant loads, analyzing the impact of noise on wireless charging power in such a complex wireless charging system is challenging. 
To perform a precise yet manageable analysis, we introduce rational simplifications based on electrical principles and the significance of components' impacts.
For this analysis, the workload is assumed to remain in a steady state, effectively modeled as a constant resistor $R_{eq}$. 
The system is segmented into three parts for sequential analysis of interference's impacts. 
Part 1 (Figure~\ref{fig:WPT_part1}) examines the impact of the changes of $v_{ad}$ at the power adapter's output on $v_{bus}$, the DC input of the inverter.
Part 2 (Figure~\ref{fig:WPT_part2}) explores how $v_{bus}$ impacts the AC voltage $v_{tx}$ across the resonant capacitor $C_p$ and TX coil at the output of the inverter.
Part 3 (Figure~\ref{fig:WPT_part3}) models the influence of the inverter's output AC voltage $v_{tx}$ on the current $i_{tx}$ in the TX coil, which directly reflects the power signal's property.

\noindent\textbf{Part 1: Transfer function from the adapter to the charger}
The influence of power adapter output voltage $v_{ad}$ on bus voltage $v_{bus}$ can be analyzed based on the model in Figure~\ref{fig:WPT_part1}.
The bus voltage $v_{bus}$ that drives the inverter is a function of the power adapter's Thevenin equivalent output voltage source $v_{ad}$, Thevenin equivalent impedance $Z_{ad}$, cable resistance $R_{cable}$, bus decoupling capacitor $C_{bus}$, and the equivalent load resistance $R_{eq}$.
Given the interfered power adapter's output voltage $v_{ad}$ in Equation~\ref{eq:setup},
the disrupted voltage $v_{bus}$ can be derived from Figure~\ref{fig:WPT_part1} as  in Equation~\ref{eq:VBUS}~\footnote{In the equations presented in this paper, we use ``$|x|$'' to represent the magnitude of a complex number $x$.}.
In Equation~\ref{eq:VBUS}, $v_{bus}$ is composed of a periodic noise with frequency $f_i$ and amplitude $K m_i V_{bus}$ superimposing on a DC component $V_{bus}$. 
$K$ is a voltage scaling factor dependent on the impedance of the model in Figure~\ref{fig:WPT_part1}.
\begin{equation}
    \label{eq:VBUS}
    \begin{split}
        &v_{bus}(t)= V_{bus}(1 + K m_i\sin(2\pi f_i t))\\
        &
        \begin{split}
            &V_{bus} = \frac{R_{eq}}{R_{eq}+R_{cable}+Z_{ad}} V_{ad}\\
            &K =  \frac{R_{eq} + R_{cable}}{|R_{eq} + R_{cable}+Z_{ad}+j2\pi f_i R_{eq}(R_{cable}+Z_{ad})C_{bus}|}
        \end{split}
    \end{split}
\end{equation}

\noindent\textbf{Part 2: Transfer function from the charger to the resonant tank} The circuit of the inverter is shown in Figure~\ref{fig:WPT_part2}. The inverter's
primary role is to convert $v_{bus}$ into AC voltage $v_{tx}$ across the resonant capacitor $C_p$ and TX coil, thereby creating the alternating magnetic field from the TX coil for power
transmission. The inverter's operation is controlled by the MCU through two
parameters: the pulse width modulation (PWM) signal with duty cycle $D$, and the power signal frequency,
$f_p$. The output of the inverter is a staircase waveform as shown in the Appendix~\ref{appx:inverter_signal}. It is fed into the resonance tank, $C_p$ and the TX coil.
The harmonics of the staircase waveform outside of the bandwidth of the resonant tank are filtered out, leaving a sinusoidal signal with a frequency equal or close to the resonant frequency of the tank. As such,
the output voltage of the inverter, $v_{tx}$, is derived in
Equation~\ref{eq:VTX}, with the derivation process detailed in
Appendix~\ref{appx:inverter_signal}. With steady-state workload, the
primary factor influencing $v_{tx}$ is $v_{bus}$, which determines the amplitude
of $v_{tx}$.
\begin{equation}\label{eq:VTX}
    v_{tx}(t) = \frac{4}{\pi}\sin{(\frac{\pi}{2}D)} v_{bus} \sin(2\pi f_p t)
\end{equation}

\noindent\textbf{Part 3: Wireless Power Transfer} The wireless power
transfer section in Figure~\ref{fig:WPT_Schematic} can be modeled in Figure~\ref{fig:WPT_part3}. 
The $v_{tx}$
drives the TX coil, generating an alternating magnetic field and transferring power
to the receiver. 
Based on the model, the current $i_{tx}$ in
the TX coil can be calculated in  Equation\ref{eq:ITX}~\footnote{In this
paper, the "$\parallel$" symbol denotes the equivalent impedance of two
parallel-connected components.}. The equivalent impedance $Z_{total}$ is a function of the
load, coupling conditions, and power signal frequency. Given that the load,
coupling conditions, and power signal frequency remain constant during this
analysis, $v_{tx}$ is the primary influential factor of the TX coil current.
\begin{equation}
    \label{eq:ITX}
    \begin{split}
        i_{tx} = \frac{v_{tx}}{Z_{total}}\text{, }Z_{total} = {(Z_{load}+Z_{rs})\parallel(j2\pi f_p)M+Z_{rp}}\\
        \text{where }
        \begin{split}
            Z_{rp} =&\frac{1}{j2\pi f_p C_p}+j2\pi f_p\cdot (L_p-M)\\
            Z_{rs} =& \frac{1}{j2\pi f_p C_s}+j2\pi f_p\cdot (L_s-M)
        \end{split}
    \end{split}
\end{equation}
\noindent\textbf{Analysis Results}
From Equations~\ref{eq:VBUS},\ref{eq:VTX}, and \ref{eq:ITX}, the TX coil current, $i_{tx}$, can be derived in Equation~\ref{eq:ITX_SUM}. 
From Equation~\ref{eq:ITX_SUM}, the schemed voltage noise on $v_{ad}$ in Equation~\ref{eq:setup} impacts $i_{tx}$ in the TX coil by modulating its amplitude. 
Because the $Z_{total}$ is a complex number, a phase difference $\phi_{total}$ exists between $i_{tx}$ and $v_{tx}$. 
The carrier signal amplitude $I_{tx}$ is determined by duty cycle D. The modulation depth $m$ is proportional to the interference depth $m_i$ and the voltage scaling factor $K$. 
\begin{equation}\label{eq:ITX_SUM}
    \begin{split}
        i_{tx}(t) =& I_{tx}(1 +  m\sin(2\pi f_i t))\sin(2\pi f_p t + \phi_{total})\\
        \text{where } & I_{tx} = \frac{4V_{bus}\sin{(\frac{\pi}{2}D)}}{\pi |Z_{total}|},
        m = K m_i\\
    \end{split}
\end{equation}
In Equation~\ref{eq:VBUS}, $K$ can be approximately estimated using typical values of $R_{eq}$(5$\Omega$), $R_{cable}$(0.1$\Omega$), $C_{bus}$(50$\mu F$).
For the interference frequencies at 1 kHz, 10 kHz, and 100 kHz, the estimated voltage scaling factor $K$ are 0.99, 0.95, and 0.30.

\noindent\textbf{Conclusion}
Existing wireless charging systems effectively attenuate high-frequency interference but are less effective against low-frequency interference. 
Therefore, low-frequency interference from the power adapter can easily propagate to the TX coil and modulate the power signal's amplitude with a modulation depth close to the interference depth.

\subsection{
Load-to-Adapter Propagation
}\label{sec:backprop}
\begin{figure}[htbp]
    \centering
    \includegraphics[width=.65\columnwidth]{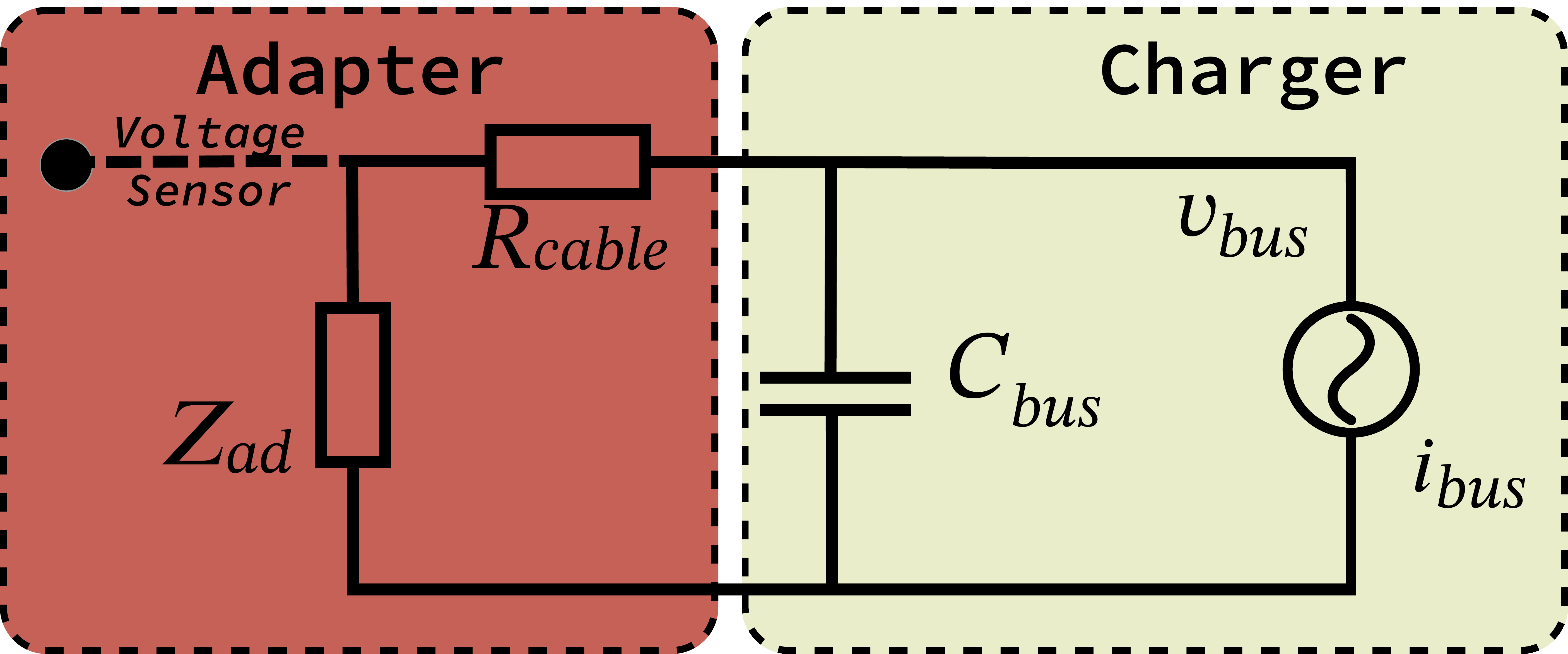}
    \caption{
    Circuit model used to analyze the impact of workload on the adapter's output voltage noise
    }
    \label{fig:ExplainFSK}
\end{figure}

An ideal power adapter is supposed to provide a constant DC voltage with minimal fluctuation, regardless of the workload behaviors.
However, a real-world power adapter's output is inevitably affected by workload behaviors mainly due to the limitations of switching regulator's close-loop bandwidth and phase margin.
This section analyzes specific workload behaviors that lead to measurable information leaks in the power adapter's output based on the circuit model shown in Figure~\ref{fig:ExplainFSK}.

The impact of workload behavior on the power adapter's output voltage noise can be analyzed by modeling the workload as an equivalent load current source $i_{bus}$ in parallel with an equivalent impedance based on the Norton's Theorem. Since this impedance is much bigger than that of $C_{bus}$, it is ignored in Figure~\ref{fig:ExplainFSK}. 
Based on the analysis in Section~\ref{sec:fwdprop}, $i_{bus}$ can be derived using $v_{bus}$, $v_{tx}$, and $i_{tx}$ per Equation~\ref{eq:BusCurrent}. 
It is composed of a DC component $I_{bus,dc}$ and an AC current $I_{bus,ac}$, which has a frequency of $2f_p$ with an amplitude proportional to $I_{bus,dc}$.
\begin{equation}\label{eq:BusCurrent}
\begin{split}
    i_{bus}(t) = \frac{v_{tx}i_{tx}}{V_{bus}} = I_{bus,dc} + I_{bus,ac} cos(4\pi f_pt + \phi_{total})\\
    I_{bus,dc} = \frac{2 I_{tx}sin(\frac{\pi}{2}D)cos\phi_{total}}{\pi}, I_{bus,ac} = \frac{I_{bus,dc}}{cos\phi_{total}}\\
\end{split}
\end{equation}

In Equation~\ref{eq:BusCurrent}, $I_{bus,dc}$ is a function of time. It is almost constant within one switching period of the inverter but varies as the load current $i_{tx}$ changes, which has much lower frequencies than that of the inverter. In a wireless charging system, we identify two workload behaviors that cause measurable signals on the output of the adapter. 
The first one is the AC current caused by the inverter's switching behaviors at the frequency of $2f_p$. 
The other is the abrupt load-change behavior. 
These behaviors are analyzed individually to understand their specific impacts on the power adapter's output voltage.

\noindent\textbf{Inverter-switching Induced Signal}
According to Equation~\ref{eq:BusCurrent}, an AC component of frequency $2f_p$ is present in the bus current, where $f_p$ is the power signal frequency controlled by the charger's MCU, typically around 140 kHz.
The voltage changes at the output of the power adapter, denoted as $\Delta V_{ad}$, can be expressed as Equation~\ref{eq:dVfsk}.
With typical values of $I_{bus,dc}$(1A), $f_p$(140 kHz), $Z_{ad}$(10 m$\Omega$), $C_{bus}$(50 $\mu$F), $R_{cable}$ (0.1 $\Omega$), and $\phi_{total}$ (70$^\circ$), the amplitude of $\Delta V_{ad}$ can be estimated as $\sim$ 10 mV.
\begin{equation}
    \label{eq:dVfsk}
    \Delta V_{ad}(t) =\frac{Z_{ad} I_{bus,dc} \cos{(4\pi f_p t + \phi_{total})}}{cos\phi_{total} (1+j4\pi f_p C_{bus}(R_{cable}+Z_{ad}))}
\end{equation}

\noindent\textbf{Load-change Induced Signal}
Based on Equation~\ref{eq:BusCurrent}, a load change, in other words, 
a change in $i_{tx}$,
also leads to the change of the load current $I_{bus,dc}$ in Figure~\ref{fig:ExplainFSK}. From  Equation~\ref{eq:dVfsk}, the load change will lead to the voltage change $\Delta V_{ad}$ at the output of the power adapter. Because of this, the load changes are detectable from $\Delta V_{ad}$. 
But as the power adapter tends to minimize $Z_{ad}$ with its high feedback control loop gain at low frequencies, the low-frequency spectrum of the $\Delta V_{ad}$ is attenuated. Only the high-frequency spectrum of the $\Delta V_{ad}$ due to the change of $i_{bus}$ remain. As a result, for an abrupt load change, which is characterized with high high-frequency spectrum, the transient voltage deviated from the nominal voltage will be observed in the output voltage, and it will rapidly settle down to its steady state value due to the adapter's close-loop feedback control. 
This results in a series of pulse signals including the load information.
This effect can be approximated as the effect of a convolution filter $\delta'(t)$. 
For a typical design, these pulses usually have small amplitudes, so they do not interfere with the normal operation of the power adapter.

\noindent\textbf{Conclusion} 
Voltage at the output of a power adapter contains the following workload behavior information signals:
the timing of load change and the frequency at which the wireless power is transferred.
Since $\Delta V_{ad}$ has a small amplitude it does not affect the functionality of a power adapter.
The signals in $\Delta V_{ad}$ are also partially masked by other voltage noise, making them not immediately distinguishable in the raw data.
However, understanding the generation and characteristics of these signals enables us to develop specialized signal processing techniques. 
These techniques can exploit the signals' unique features to successfully extract the embedded information.

%% file: files/implementation.tex
\section{Preliminary Attack Vectors}
\label{sec:attack_vector}

Through comprehensive analysis, the two questions raised in
Section~\ref{sec:vulnerability_analysis} have been answered, yielding two
essential insights concerning a wireless charging system:

\begin{tcolorbox}[
    skin=widget,
    left=0mm,
    right=0mm,
    top=0mm,
    bottom=0mm,
    boxrule=0.75mm,
    colback=InsightLight,
    colframe=Insight,
    width=(\linewidth),
    segmentation style={solid, line width=0.35mm, dashed},
    title=\hliwona{Insights},
    ]

    \hliwona{Insight 1: The manipulated low-frequency signals at the output of the power adapter can propagate to
    the TX coil and modulate the power signal with limited attenuation and
    distortions.}  
    
    \tcbline
    
    \hliwona{Insight 2: Information such as frequency, timing and amplitude of both the inverter switching and charging load change is reflected by the voltage noise at the output of the power adapter.}  
        
\end{tcolorbox}

This section showcases three practical attacks derived from our insights. We cover exploiting voice signal induction in charging smartphones (Section~\ref{sec:attk_vec_voice}), injecting malicious Qi messages to alter charging control (Section~\ref{sec:attk_vec_QiInterfere}), and recovering communication messages through voltage noise analysis (Section~\ref{sec:attk_vec_QiEavesdrop}).

\subsection{Attack Vector 1: Voice Injection}\label{sec:attk_vec_voice}

This section introduces our first practical attack vector, which is injecting
voice signals into the charged smartphones. The most significant information in
typical voice signals is in the frequency band below 10 kHz~\cite{monson2014perceptual}. Therefore,
according to \hliwona{Insight 1}, when a voice signal is added to the power
adapter's output voltage, it can modulate the power signal at the TX coil with limited
attenuation and distortions. A recent study~\cite{Dai:2023:SP} has demonstrated that an
AM-modulated magnetic field can cause magnetic-induced sound (MIS) in the
microphone circuits of modern smartphones through magnetic couplings. Thus, by
adding voice signals to the power adapter's output, we will be able to inject
voice signals into the charged smartphones exposed to this intense magnetic
field. To validate this sound-inducing mechanism, we conducted tests on
an iPhone SE and a Pixel 3 XL with a Renesas P9242-R-EVK wireless charger. In
these tests, we recorded the activation commands of these two smartphone
assistants spoken by their owners. When the iPhone SE is under charging, the waveform
of ``Hey Siri'' is added to the supply voltage, and a recording application on
the smartphone is activated to capture any potential audio signals. Similarly,
for the Pixel 3 XL, the test involves adding the waveform of “Hey
Google” to the supply voltage and recording any resulting audio signals. The
recording process takes place in a normal office environment with a reasonable
level of background noise.

\begin{figure}[htbp]
    \centering
    \subfloat[Spectrograms of ``Hey Siri'']{\includegraphics[width=\columnwidth]{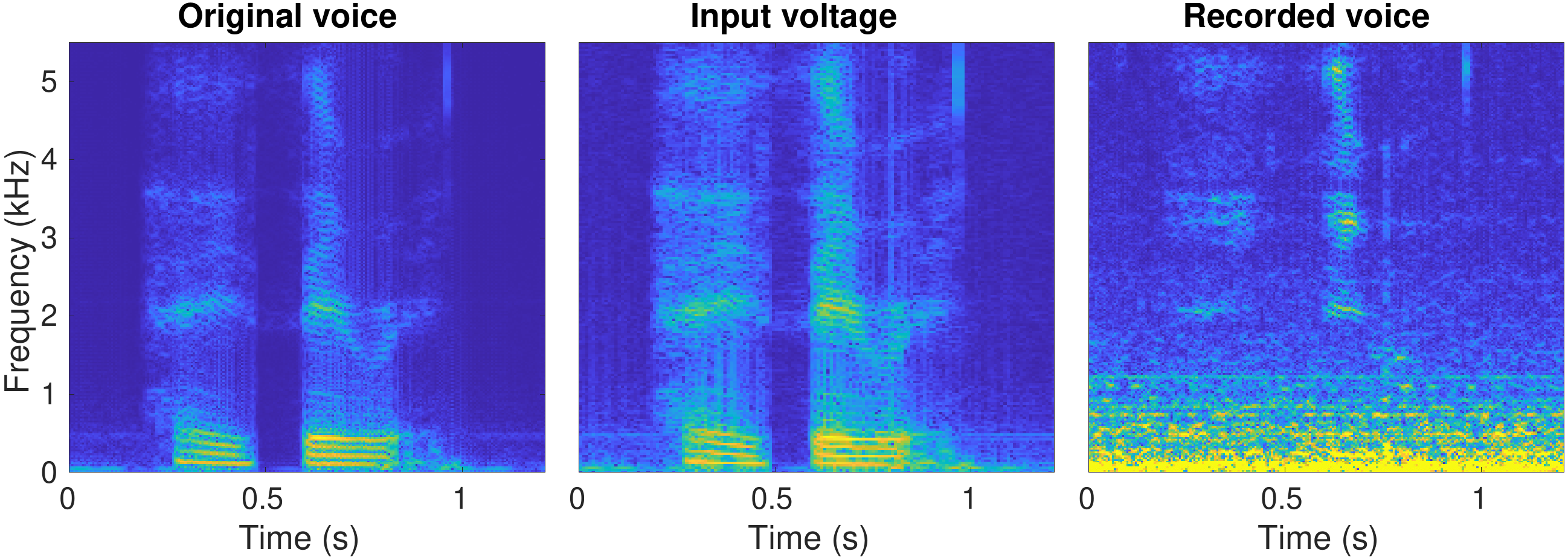}}\\
    \subfloat[Spectrograms of ``Hey Google'']{\includegraphics[width=\columnwidth]{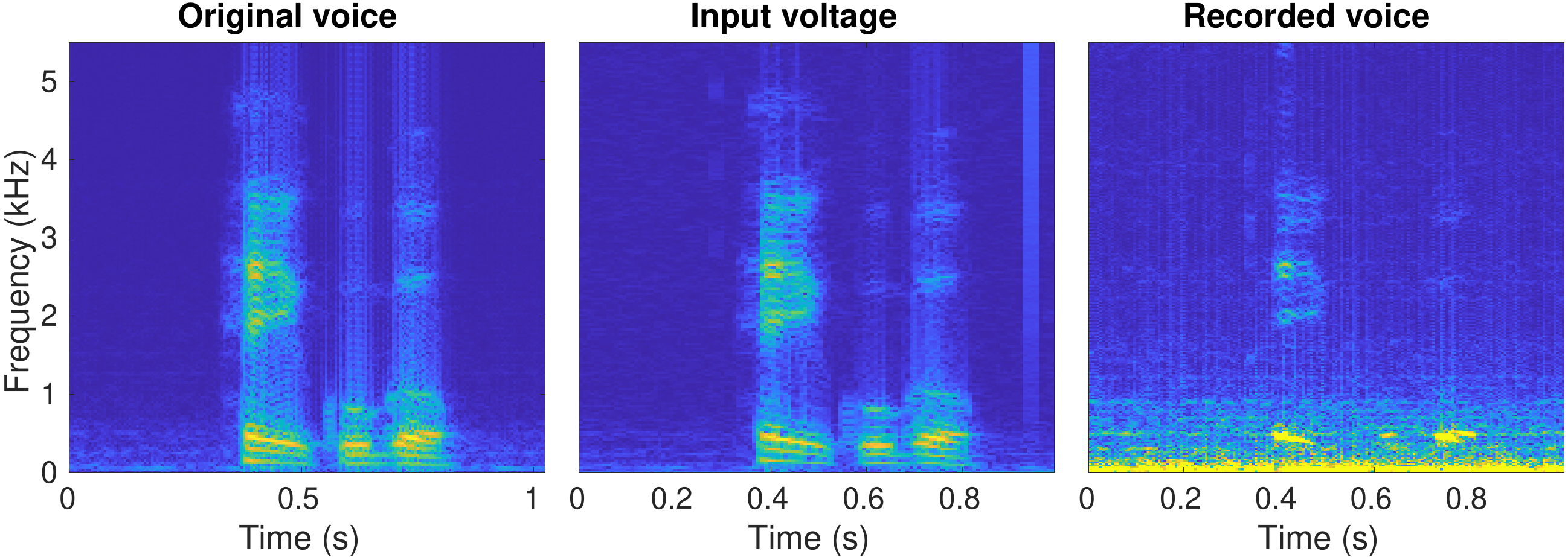}}
    \caption{Spectrograms of signals collected during injecting MIS to smartphones}
    \label{fig:hey_siri_comparison}
\end{figure}

Figure~\ref{fig:hey_siri_comparison} compares the spectrograms of the original
voice signal, the adapter's interfered output voltage signal, and the signal captured by the
microphone during charging. It is evident from the spectrograms that some
features of the original sound signal are recognizable in the MIS. However, the
signal-to-noise ratio (SNR) of the MIS is affected by a couple of key factors.
First, when the intensity of the resulting sound is weak, some patterns are
overwhelmed by background noise. To counter this, we can increase the interference
depth $m_i$ to enhance the SNR. Secondly, although the analysis in
Section~\ref{sec:vulnerability_analysis} demonstrates limited attenuation for
low-frequency signals, different frequency components of the original voice
signals are still subject to different attenuation. This unequal
attenuation across the frequency band can distort the signal waveform and result in the
loss of audio features.

A security implication of this attack vector is that an attacker may exploit this mechanism to inject voice commands and control the voice assistants in the charged smartphones. 
The voice assistants will likely recognize a considerable amount of features preserved in the MIS and execute the commands.

\subsection{Attack Vector 2: Qi Message
Injection}\label{sec:attk_vec_QiInterfere} In this section, we explore the
attack vector of injecting ASK-modulated Qi messages into the communication
channels between RX and TX devices. During charging, the RX device modulates the
power signal at a frequency of approximately 2 kHz. As per \hliwona{Insight 1},
an interference signal around this frequency at the output of the power adapter can modulate the power signal with
small attenuation. Therefore, it is feasible to inject synthesized ASK
modulation signals, which strictly adhere to Qi communication protocols, into the output of the power adapter to deceive
the TX device.

\begin{figure}[htbp]
    \centering
    \subfloat{\includegraphics[width=\columnwidth]{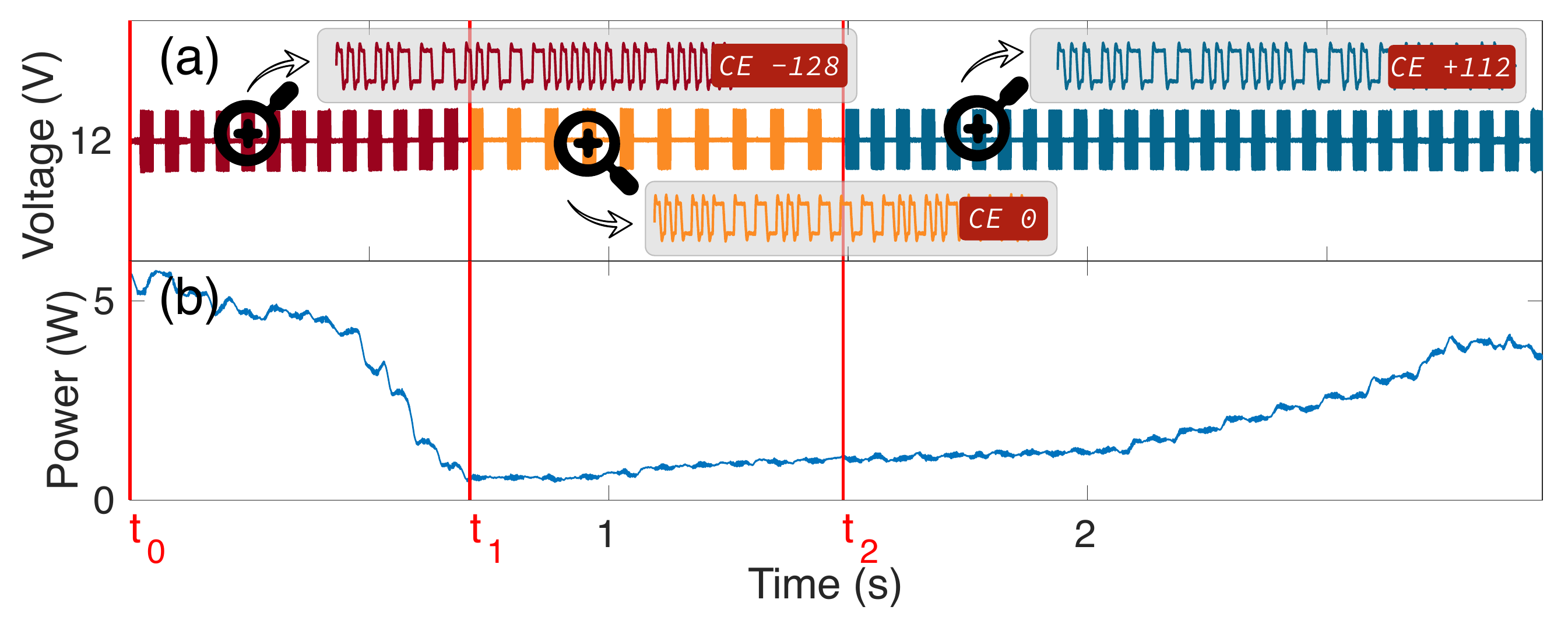}}
    \caption{Inject \texttt{CE} packets to manipulate the charging power. (a) Input voltage with injected \texttt{CE} packets. (b) Charging power affected by the injected packets}
    \label{fig:CE_Power_Manipulation}
\end{figure}

To demonstrate this capability, we used a Renesas P9242-R-EVK wireless charger
to charge an iPhone SE. 
We injected fake \texttt{CE} packets into the power adapter's output voltage to decive the charger.
The charger adjusted its charging power as directed by the fake commands. The
results are displayed in Figure~\ref{fig:CE_Power_Manipulation}, where the voltage
trace shows three different \texttt{CE} messages, \texttt{CE}(-128),
\texttt{CE}(0), and \texttt{CE}(+112), inserted at timestamps $t_0$, $t_1$, and
$t_2$, respectively. The power trace correlates the output power changes with
the respective \texttt{CE} values, confirming that the charging power was
manipulated as expected.

A security implication of this attack vector is that it provides
the attacker with a communication channel to send malicious messages to
chargers. 
Injecting interference at the ASK modulation frequency into the power adapter's output
can disrupt the genuine packets sent from RX devices and hijack the in-band communication. 
When the Qi communication is compromised, many charging
safety mechanisms that heavily rely on this communication can be invalidated as
well. An attacker can exploit this attack vector to induce hazardous
charging processes that could severely damage the charged devices.

\subsection{Attack Vector 3: Qi Message Eavesdropping}\label{sec:attk_vec_QiEavesdrop}

This section investigates the attack vector that enables an attacker to recover
Qi messages using the voltage trace measured at the power adapter's output.
As introduced in Section~\ref{sec:background}, the RX and TX devices modulate
the power signal using ASK and FSK modulations, which impact the power signal by
shifting the load and altering the power signal frequency, respectively.
According to \hliwona{Insight 2}, the load power modulation will lead to measurable
signals at the power adapter's output. However, such information may not be
directly visible in the measured raw traces due to the low intensity of these
signals. Specialized signal processing techniques that target these signal
features are necessary to extract this information. In the remaining part of
this section, we present our methodologies for processing the signal to recover
messages using ASK and FSK modulations. A voltage trace captured at the
beginning of the charging initiation process between a Renesas P9242-R-EVK
wireless charger and an iPhone SE will be used to demonstrate these methodologies.

\noindent\textbf{ASK Modulation Eavesdropping}
Analysis in Section~\ref{sec:backprop} indicates that the effect of a load transition on the charged device on the power adapter's output voltage can be represented by being filtered with a convolution filter $\delta'(t)$. 
Therefore, to recover the waveform of the ASK modulation signal, we introduce the convolution kernel $h_1(t)$ in Equation~\ref{eq:filter}. 
$h_1(t)$ is a triangle pulse smoothing filter designed to counteract the effects of the equivalent filter $\delta'(t)$. 
The combined result forms a matched filter that detects transitions between \texttt{LOW} and \texttt{HIGH} at the frequency of $f_{ASK}$. 
Given that BMC encoding schemes are used for bit encoding, a significant feature for distinguishing the transmission of \texttt{ZERO} and \texttt{ONE} is the phase shift pattern of the signal at frequency $f_{ASK}$. 
Based on this characteristic, we further employ the filter $h_2(t)$ in Equation~\ref{eq:filter} to enhance such phase shift patterns for the signals with frequency $f_{ASK}$.
\begin{equation}\label{eq:filter}
    \begin{split}
        h_1(t) =& 1 - f_{ASK}|t| \text{, } -\frac{1}{f_{ASK}} \leq t \leq \frac{1}{f_{ASK}}\\
        h_2(t) =& \delta(t - \frac{1}{2f_{ASK}}) - \delta(t + \frac{1}{2f_{ASK}})
    \end{split}
\end{equation}

\begin{figure}[htbp]
    \centering
    \includegraphics[width=\columnwidth]{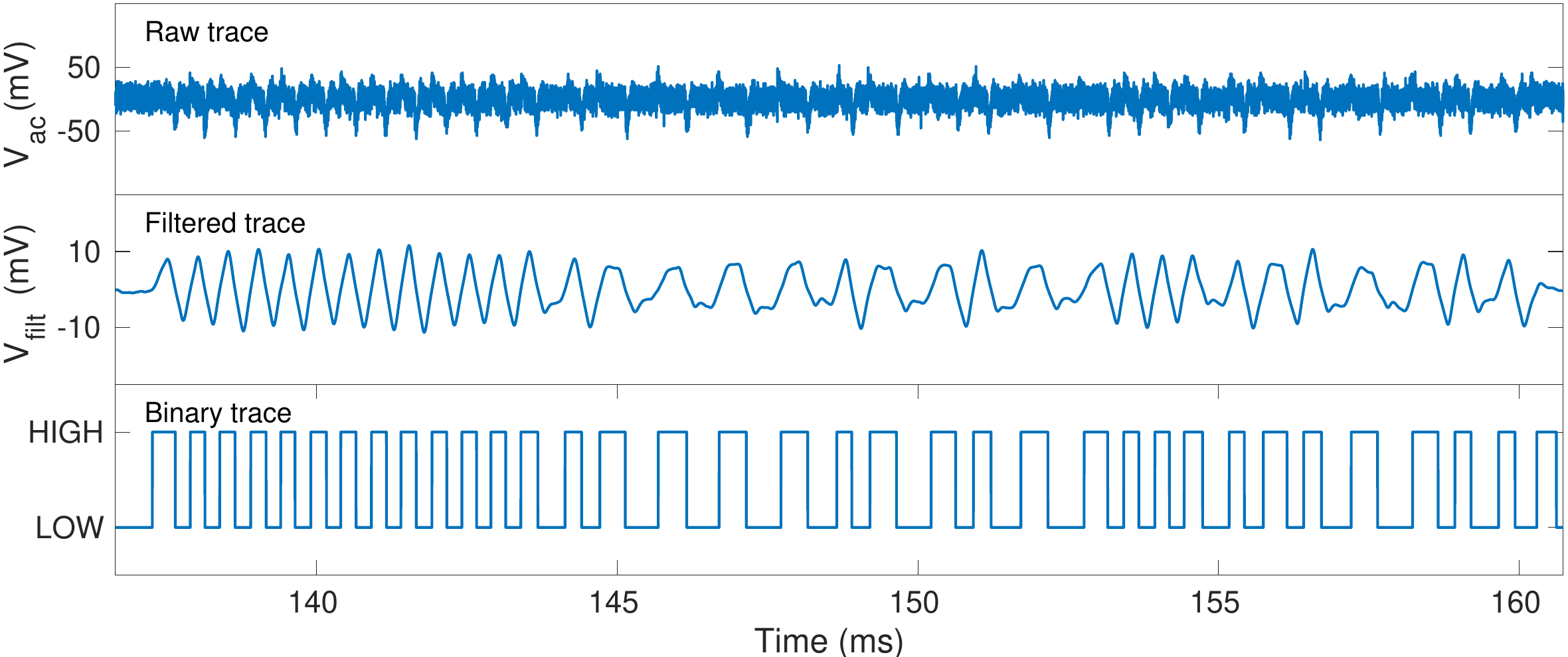}
    \caption{ASK modulation recovery}
    \label{fig:eavesdrop-ASK}
\end{figure}

The effectiveness of these filters is demonstrated in Figure~\ref{fig:eavesdrop-ASK}.
While some pulses are visible in the raw trace, the modulating pattern is unclear.
After filtering, we can effectively recover the signals with clear ASK modulation patterns, which can be further decoded into the binary \texttt{HIGH}-\texttt{LOW} sequence. 
For this specific example, we recover a \texttt{SIG} packet with the value 0x84 after decoding. Using the same technique, we can also recover other data packets sent by the power receiver, such as \texttt{ID}, \texttt{CFG}, \texttt{FOD}, \texttt{GRQ}, \texttt{SRQ}, \texttt{RP}, \texttt{CE}, etc.

\begin{figure}[htbp]
    \centering
    \includegraphics[width=\columnwidth]{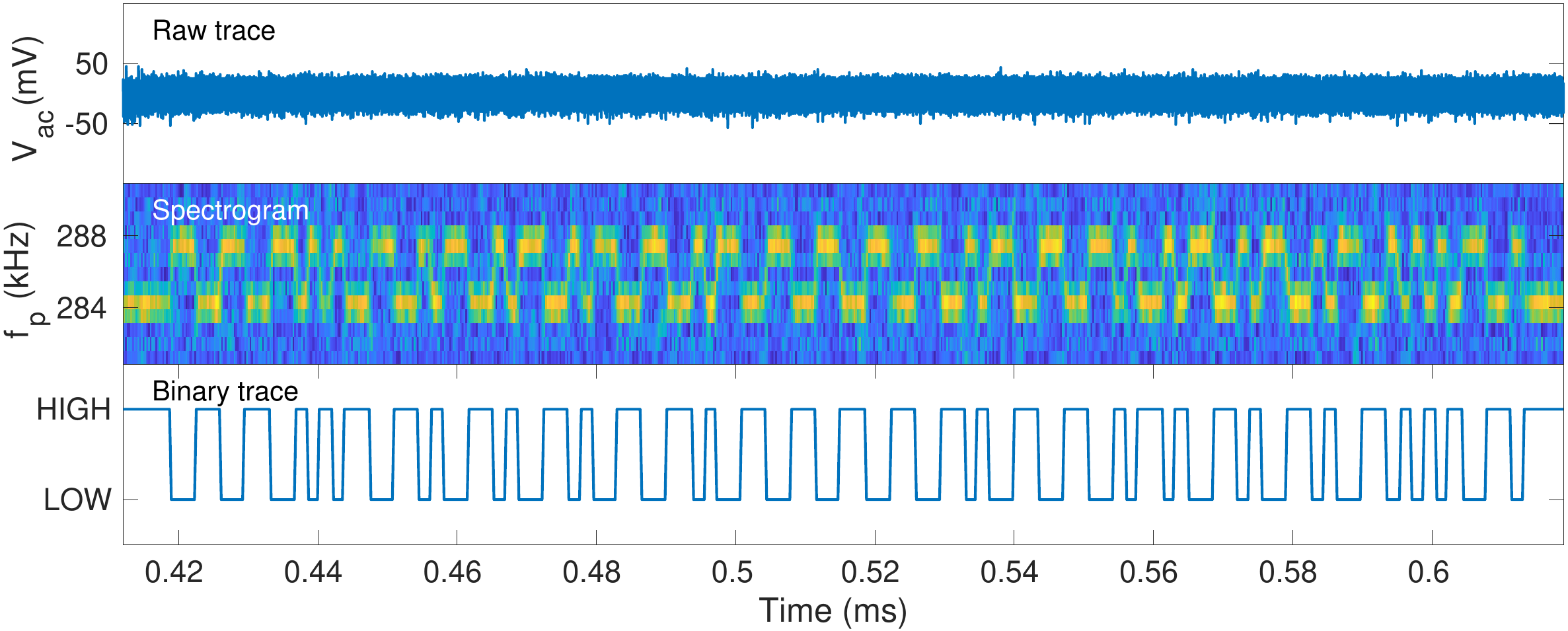}
    \caption{FSK modulation recovery}
    \label{fig:eavesdrop-FSK}
\end{figure}

\noindent\textbf{FSK Modulation Eavesdropping} Analysis in
Section~\ref{sec:backprop} indicates that a weak signal at the frequency of
$2f_p$ can be measured at the power adapter's output. With the TX device
using FSK modulation to transmit data by altering the power signal frequency
$f_p$, an attacker can track the frequency changes to recover modulation
signals. To extract these frequency-domain features, we perform a discrete
Fourier transform (DFT) on the measured raw voltage trace and analyze the spectrogram. As
the results in Figure~\ref{fig:eavesdrop-FSK} show, while no features are visible in the time domain trace, distinctive patterns exist in the
frequency domain. When $f_p$ is around 140 kHz, frequency-switching
patterns near 280 kHz are clear. In this case, we can decode the derived
binary sequence to recover an \texttt{ID} packet, which discloses the charger's
identification.

This attack vector reveals several security concerns. Initially, it exposes that normal charging processes unintentionally leak charger and device models, allowing attackers to profile and target specific devices. 
Furthermore, combining eavesdropping on and injecting Qi messages grants attackers the ability to simulate a legitimate receiving device's behavior. This deception could lead the charger to initiate power transfer under hazardous conditions, all achievable with mere access to the power adapter, indicating a significant threat to wireless charging security.

\section{Practical Attacks Implementation}

This section outlines conducting three practical attacks detailed in Section~\ref{sec:attack_vector}. It includes a setup for these attacks (Section~\ref{sec:exp_setup}), a method to manipulate voice assistants via injected commands (Section~\ref{sec:attack_voice}), a wireless power toasting attack causing charger-induced device damage (Section~\ref{sec:attack_wpt}), and a foreign object destruction attack misleading the charger to damage non-targeted objects (Section~\ref{sec:attack_fod}).

\subsection{Experimental Setup}
\label{sec:exp_setup}

\begin{figure}[ht]
    \centering
    \includegraphics[width=0.98\columnwidth]{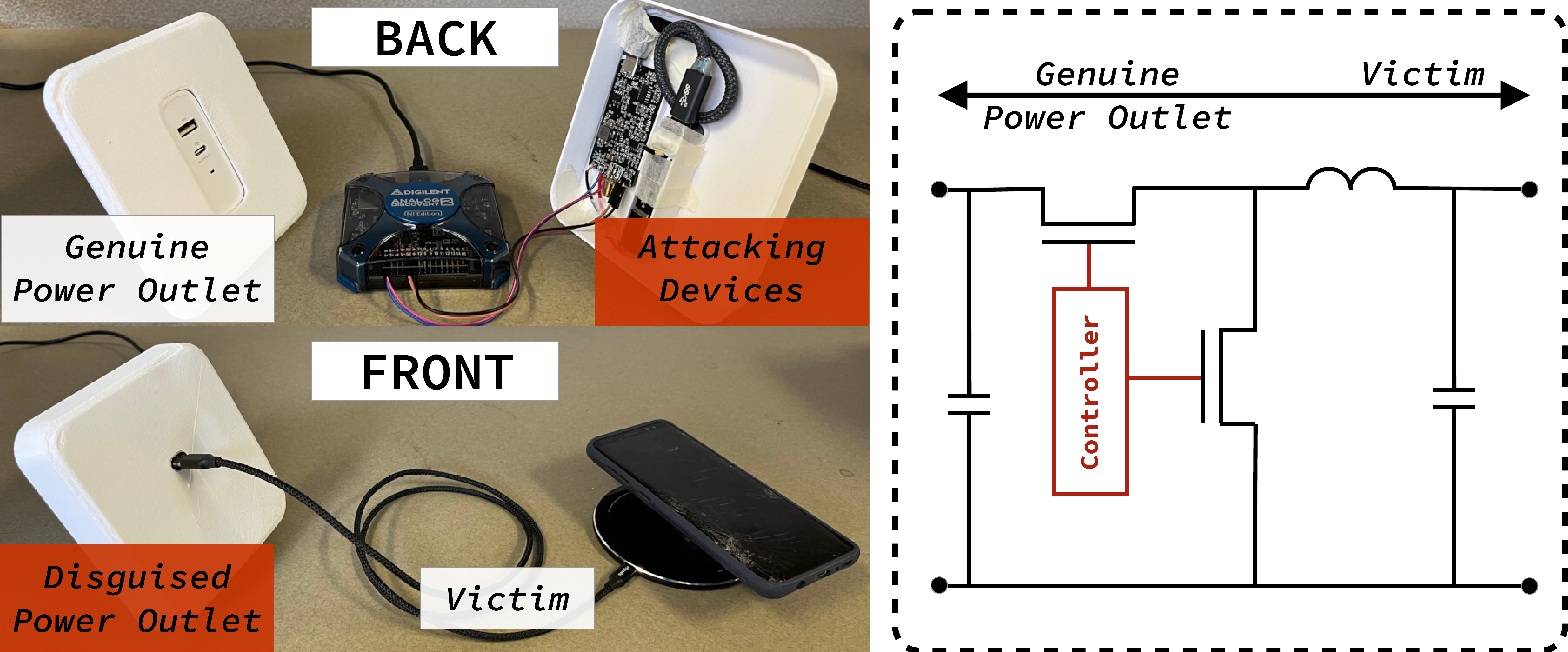}\vspace{-1em}
    \subfloat[\label{fig:setup}]{\hspace{.55\linewidth}}
    \subfloat[\label{fig:setup_sch}]{\hspace{.45\linewidth}}
    \caption{Hardware setups used to implement \VoltSchemer: 
    (a) experimental setup: 
    (b) voltage manipulator design.}%
\end{figure}

In Figure~\ref{fig:setup}, we show a practical attacking setup that can be easily found in real-life scenarios. The attacker
employs a disguised power port, which appears to be a regular USB-C port from the
front but conceals a USB-C plug at the back. Behind this facade lies an
attacker-controlled voltage manipulator connected between the power pins of the
two USB-C connectors. As illustrated in Figure~\ref{fig:setup_sch}, this
manipulator alters the switching patterns of two MOSFETs to superimpose the manipulated AC voltage fluctuations onto the DC voltage. 

In our experiment, we used the Analog Discovery 2 (AD2) as a controller to process the measured output and generate signals to control the injected noise waveform and intensity. 
For mass production, this prototype can be significantly miniaturized by substituting AD2 with a compact controller chip, akin to the size depicted in Appendix~\ref{appx:chip_size}.
Installation of this device only requires simply plugging it into a COTS power adapter's power port and replacing its functionality. 
Given the uniform function of power adapters to supply DC voltage, this method is universally applicable to all COTS power adapters. 
We tested Apple, Google, and Amazon power adapters to verify our ability to inject configurable voltage noise with specific $m_i$ and $f_i$ values. 
We show wireless chargers connected to this disguised power port are vulnerable to various attacks. 
The efficacy and practicality of \VoltSchemer\ are validated through evaluations on 9 popular wireless chargers listed in Table~\ref{tab:evaluation_devices}, featuring a range of manufacturers and power ratings.

\definecolor{5W}{HTML}{F1FAEE}
\definecolor{10W}{HTML}{669BBC}
\definecolor{15W}{HTML}{CA3402}
\begin{table}[htbp]
    \small
    \centering
    \caption{List of evaluated wireless chargers}
    {\fontfamily{ppl}\selectfont
        \begin{tabular}{|c|c|c|c|}
            \hline
            ID.              & Manufacturer & Model        & Rated Power \\
            \hline
            \rowcolor{5W} 1  & KEYOMOX      & B0835LGZ9B   & 5W          \\
            \hline
            \rowcolor{10W} 2 & Anker        & A2503        & 10W         \\
            \rowcolor{10W} 3 & COCOEYE      & Wi-II        & 10W         \\
            \rowcolor{10W} 4 & FDGAO        & B413         & 10W         \\
            \rowcolor{10W} 5 & Philips      & DLP9035BC/27 & 10W         \\
            \rowcolor{10W} 6 & YOOTECH      & F500         & 10W         \\
            \hline
            \rowcolor{15W} 7 & Renesas      & P9242-R-EVK  & 15W         \\
            \rowcolor{15W} 8 & TOZO         & W1           & 15W         \\
            \rowcolor{15W} 9 & WaiWaiBear   & PAWCS11B     & 15W         \\

            \hline
        \end{tabular}
    }
    \label{tab:evaluation_devices}
\end{table}

%% file: files/VoiceInjection.tex
\subsection{Voice Assistant Manipulation}\label{sec:attack_voice} 

As discussed in Section~\ref{sec:attk_vec_voice}, by interfering with the supply voltage of the wireless charger, voice signals can be induced in the microphone of a charged smartphone. 
This section shows how this method can be used to manipulate voice assistants, which are widely used in modern smartphones. 
To assess the practical impact of this voice assistant manipulation attack, we focus on two key aspects. 
First, we measure the maximum distance between the charger and the smartphone at which the attack remains effective. 
Additionally, to confirm the attack's versatility in controlling voice assistants, we test it with a range of commonly used voice commands.

\subsubsection{Attack Evaluations}

We evaluated nine COTS wireless chargers, as listed in Table~\ref{tab:evaluation_devices}, using two smartphones: the iPhone SE and the Pixel 3 XL. 
The iPhone SE, manufactured by Apple, utilizes the iOS system and employs Siri as its voice assistant. 
The Pixel 3 XL, manufactured by Google, operates on the Android system and employs Google Assistant. 
Leveraging \textbf{Attack Vector 3}, the manufacturer information of the targeted smartphone can be procured from the eavesdropped \texttt{ID} packet sent by it.

\noindent\textbf{Evaluations of Attacking Distance}
Because Qi Wireless charging requires precise alignment between TX and RX coils for stable power transfer, the maximum measurable attacking distance is limited to $\sim$ 3 cm.
Beyond this distance, the charging process is terminated. 
To facilitate evaluations of longer attack distances, we placed a Renesas P9221-R power receiver on the charging pad to keep the wireless charger running even when the smartphone is moved out of the charging range, ensuring consistent power transfer during the evaluation. 
We introduced interference using the voice assistant activation commands ``Hey Siri'' and ``Hey Google'' to target the voice assistants of the iPhone SE and Pixel 3 XL, respectively.
The interference depth is fixed at 0.3, which is the minimal level sufficient to activate all voice assistants without disrupting power transfer. 
We measured the maximum distances at which voice assistants can be successfully activated by placing the smartphone at different distances from the charging pads.

\begin{figure}[htbp]
    \centering
    \includegraphics[width=\columnwidth]{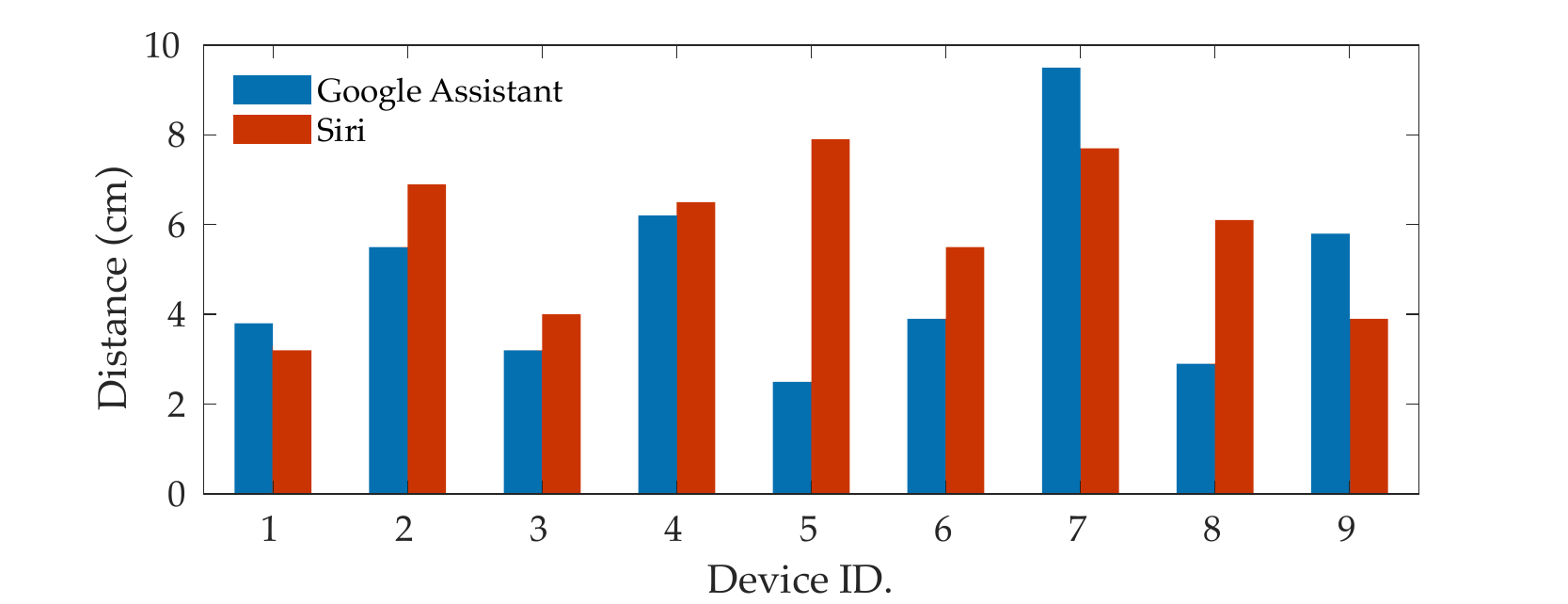}
    \caption{Maximum attacking distance}
    \label{fig:voice_distance}
\end{figure}

The evaluation results in Figure \ref{fig:voice_distance} indicate that although successful attacks have different maximum attacking distances from 3 cm to 10 cm between the chargers and the smartphones for different wireless chargers, the maximum distance is not smaller than the 3 cm wireless charging range limited by the misalignment constraint in Qi standard, therefore, the voice assistant manipulation attacks can always be successfully conducted to the charged smartphones.

\noindent\textbf{Evaluations of Voice Commands}
We evaluated six frequently used voice commands on the iPhone SE and Pixel 3 XL to assess the effectiveness of injecting different voice commands across various wireless chargers and smartphones.
These commands are designed to prompt specific actions with the voice assistant, including activating the assistant, initiating a phone call, browsing a website, launching an app, using the speaker, and controlling the camera. 
The system's resilience to a voice assistant manipulation attack depends on many factors, including the electrical characteristics of the system, the features of the voice signals, and the algorithms of the voice assistants.
To launch a successful attack on a more resilient system, a higher interference depth $m_i$ is required to induce a stronger voice signal.
Meanwhile, an excessively high interference depth $m_i$ may intermittently disrupt the charging process and compromise the stealthiness of the attack.
For instance, we observed that intermittent charging interruptions start occurring when $m_i$ exceeds 0.35 and become more frequent when $m_i$ surpasses 0.5. 
Therefore, our evaluations aim to identify the minimum interference depth $m_i$ required for successful command injection. Lower $m_i$ means more efficient and stealthier attacks. 
We increased the interference depth by a 0.005 step from 0 to measure this threshold precisely.

\begin{figure}[htbp]
    \centering
    \includegraphics[width=\columnwidth]{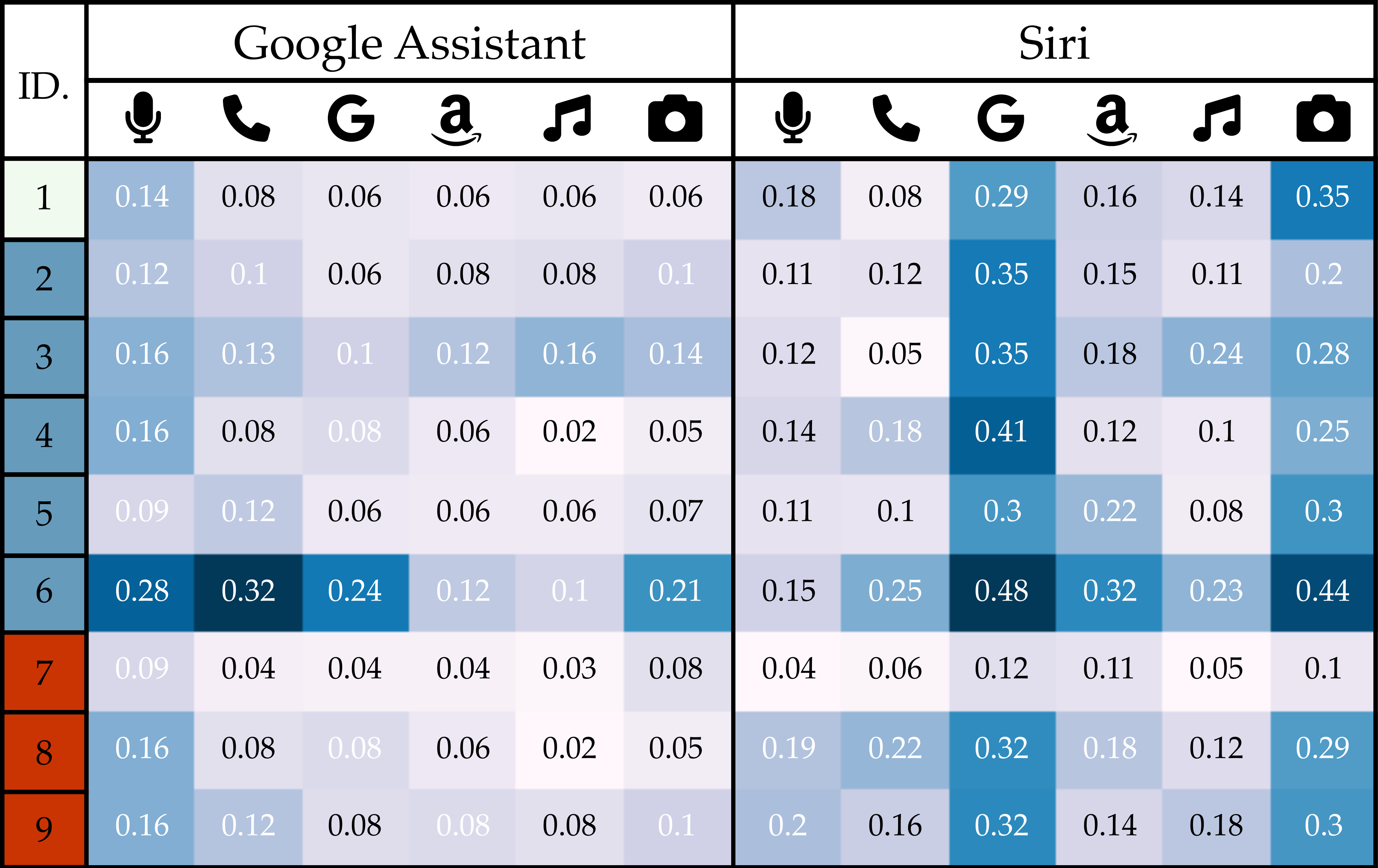}\\
    \faMicrophone: Hey Siri/Google, \faPhone: Call Alice, \faGoogle: Go to google.com, \\
    \faAmazon: Open Amazon, \faMusic: Play music, \faCamera: Take a selfie
    \caption{Required interference depth of successful command injection to Siri (iPhone) and Google Assistant (Pixel)}
    \label{fig:voiceInjection}
\end{figure}

The results in Figure~\ref{fig:voiceInjection} demonstrate how effective this attack is on various devices and voice commands. 
105 of 108 voice commands can be successfully injected at interference levels lower than 0.35. Only 3 of 108 injections require an interference depth between 0.35 and 5. This shows the efficacy and feasibility of our voice assistant manipulation attacks.

%% file: files/charging_manipulation.tex
\subsection{Wireless Power Toasting}\label{sec:attack_wpt}

As demonstrated in Section~\ref{sec:attk_vec_QiInterfere}, injecting
interference with ASK modulation patterns into the supply voltage enables an
attacker to manipulate the charging control. This section illustrates how this
capability can be used to launch a wireless power toasting attack, potentially
damaging the charged smartphones through overcharging and overheating. Vendor
documentation indicates that modern smartphones typically incorporate multiple
techniques to mitigate risks associated with overcharging and
overheating~\cite{apple2023support,google2023support}. Therefore, a strategic
approach is necessary to circumvent these protection measures. Smartphones
typically adopt three protection measures: \emph{P1} - terminating charging,
\emph{P2} - shutting down all apps and disabling user interaction, and \emph{P3}
- initiating an emergency shutdown. While \emph{P2} and \emph{P3} focus on
reducing heat generation within the smartphone itself, \emph{P1} poses a direct
challenge to the attack. This protection involves two actions: commanding the
charger to stop power transmission by sending an \texttt{EPT} message and
deactivating the smartphone's power receiving module. The charger may cease
power transmission either immediately upon receiving an \texttt{EPT} message or,
alternatively, due to a loss of communication if it fails to receive regular
\texttt{CE} and \texttt{RP} packets from the smartphone.

Thus, besides increasing charging power with \texttt{CE} packets, we developed a
strategy fulfilling two additional critical requirements to execute the wireless
power toasting attack: \Circled{\textbf{1}} Inject interference to disrupt
legitimate Qi messages from the smartphone to prevent charging termination
triggered by \texttt{EPT} packets. \Circled{\textbf{2}} Continuously inject
\texttt{CE} and \texttt{RP} packets regularly to sustain the Qi communication
with a charger, even after the smartphone's power receiving module is
deactivated.

\subsubsection{Attack Evaluations}

\begin{figure}[htbp]
    \centering
    \includegraphics[height=.4\columnwidth]{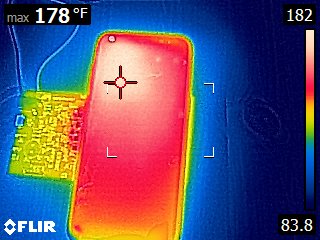}
    \caption{Thermal image of the overheated phone}
    \label{fig:phone_heatmap}

\end{figure}

To evaluate whether the wireless power-toasting attack can succeed even with the protection measures employed in smartphones, we conducted experiments using a Samsung Galaxy S8 smartphone~\footnote{A different smartphone was used for potentially destructive experiments.}. 
Upon injecting \texttt{CE} packets to increase power, the temperature rapidly rose. 
Shortly after, the phone tried to halt power transfer (\emph{P1}) by transmitting \texttt{EPT} packets due to overheating, but the voltage interference introduced by our voltage manipulator corrupted these, making the charger unresponsive. 
Misled by false \texttt{CE} and \texttt{RP} packets, the charger kept transferring power, further raising the temperature.
The phone further activated more protective measures: closing apps and limiting user
interaction (\emph{P2}) at 126 F$^\circ$ and initiating emergency shutdown (\emph{P3}) at 170 F$^\circ$.
Still, power transfer continued, maintaining a dangerously high temperature,
stabilizing at 178 F$^\circ$ as per Figure~\ref{fig:phone_heatmap}. The actual
core temperature inside the phone often surpasses the surface temperature.

In experiments conducted on all evaluated chargers, we recorded the maximum
charging power and highest temperature each charger could induce on a
smartphone, and checked the activation of three thermal protection measures,
\emph{P1}, \emph{P2}, and \emph{P3}. Using a thermal camera and battery health
monitor app, we monitored the surface and core battery temperatures on the
phone. The measured core temperature using the app stopped at 131 F$^\circ$ due to the
activation of \emph{P2}, although the actual temperature continuously increased far beyond that. The recorded surface temperature with the thermal camera reaches as high as 179 F$^\circ$. As detailed in the results from
Table~\ref{tab:evaluation_charging}, our results reveal concerning findings. All
compromised chargers pushed the phone's temperature beyond its specified working
temperature (95F$^\circ$). High-power chargers caused even more thermal
stress. All tested chargers, when compromised, can trigger the power receiving
termination protection measure. High power chargers (\textasciitilde10W) can
force the phone into the second thermal protection mode, restricting user
interactions. In the worst scenarios, \textasciitilde15W chargers can force
smartphones to shutdown due to excessive heat. Such persistent overheating
attack presents a much higher risk than typical phone-generated overheating,
potentially causing battery failure or explosion.

\begin{table}[htbp]
    \centering
    \small
    \caption{Charging Power Manipulation Range}
    {\fontfamily{ppl}\selectfont
        \begin{tabular}{|c|c|c|c|c|c|c|}
            \hline
            ID.              & \emph{P1}     & \emph{P2}     & \emph{P3}     & \makecell{Core Temp              \\ ($^{\circ}$F)} & \makecell{Surf Temp \\ ($^{\circ}$F)} & \makecell{PWR \\ (W)} \\
            \hline
            \rowcolor{5W} 1  & \cmark & \cmark & \xmark & 131+                & 124 & 9    \\
            \rowcolor{10W} 2 & \cmark & \xmark & \xmark & 109.4               & 109 & 5.2  \\
            \rowcolor{10W} 3 & \cmark & \xmark & \xmark & 125.42              & 118 & 7.3  \\
            \rowcolor{10W} 4 & \cmark & \cmark & \xmark & 131+                & 125 & 9.3  \\
            \rowcolor{10W} 5 & \cmark & \cmark & \xmark & 131+                & 126 & 7.6  \\
            \rowcolor{10W} 6 & \cmark & \cmark & \xmark & 131+                & 126 & 9.2  \\
            \rowcolor{15W} 7 & \cmark & \cmark & \cmark & 131+                & 179 & 18   \\
            \rowcolor{15W} 8 & \cmark & \cmark & \cmark & 131+                & 173 & 17   \\
            \rowcolor{15W} 9 & \cmark & \cmark & \xmark & 131+                & 149 & 13.2 \\
            \hline
        \end{tabular}
    }
    \label{tab:evaluation_charging}
\end{table}

%% file: files/foreign_object_destruction.tex
\subsection{Foreign Object Destruction}\label{sec:attack_fod}

Leveraging \textbf{Attack Vector 2} and \textbf{Attack Vector 3}, an attacker
can inject and receive Qi communication packets, thus enabling interactive
communication with the wireless charger and mimicking a legitimate RX device.
This capability allows an attacker to manipulate the charger into transferring
power even without actual RX devices present. This section demonstrates the
foreign object destruction attack, where the charger is controlled to damage
foreign objects by transferring power to them and causing excessively high
temperatures.

\begin{figure}[htbp]
    \centering
    \subfloat[Essential communications for initiating power transfer\label{fig:initiate_approaches}]{\includegraphics[width=\columnwidth]{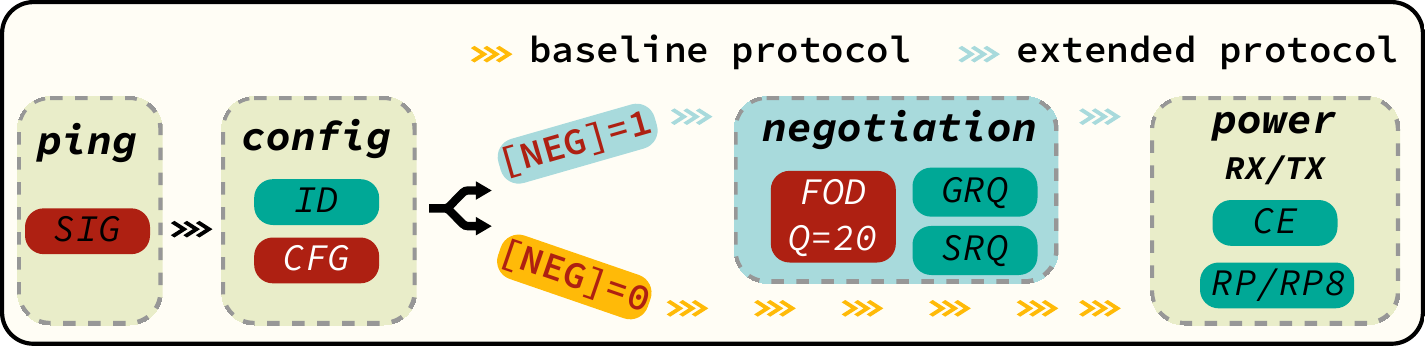}}\\
    \subfloat[Interfered voltage trace and measured output power\label{fig:power_transfer_initiate}]{\includegraphics[width=\columnwidth]{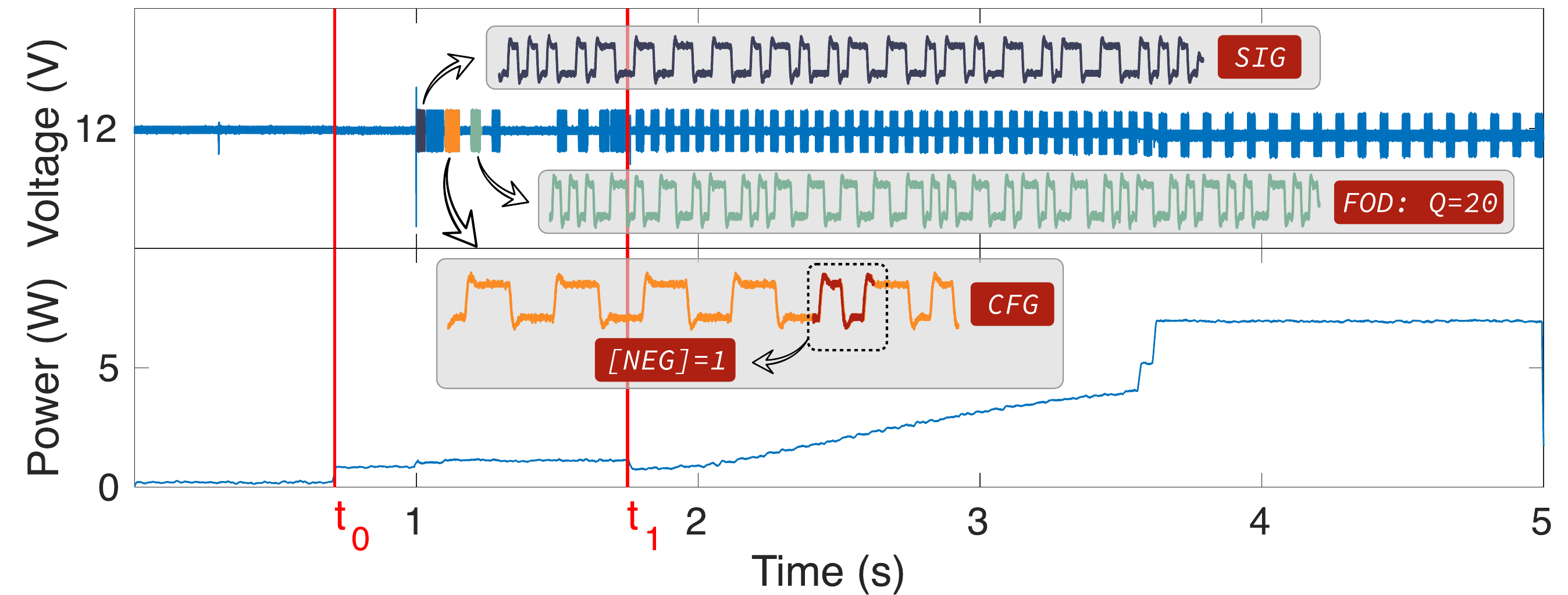}}
    \caption{Process of initiating power transfer}
\end{figure}

Through an in-depth analysis of the Qi wireless charging protocol, we identified
critical steps to initiate power transfer to foreign objects. The procedure is
detailed in Figure~\ref{fig:initiate_approaches}, and its practical
implementation is demonstrated in Figure~\ref{fig:power_transfer_initiate},
which shows the interfered voltage and output power traces during the
manipulation of a charger to transmit power to a metal foil. The process
involves three key stages: \texttt{\textbf{ping}},
\texttt{\textbf{configuration}}, and \texttt{\textbf{negotiation}}. In the
\texttt{\textbf{ping}} stage, starting at $t_0$, the charger applies a power
signal and awaits a response. We must respond with a \texttt{SIG} packet within
the required timeframe to proceed to the \texttt{\textbf{configuration}} stage.
Here, a fabricated device ID is sent to the charger, and the power protocol is
selected by setting the \texttt{NEG} bit in the \texttt{CFG} packet. To ensure
higher charging power, the extended protocol is selected by setting \texttt{NEG}
to 1 and proceeding to the \texttt{\textbf{negotiation}} stage. Otherwise, the
charger defaults to the baseline protocol with a maximum charging power of 5W.
During \texttt{\textbf{negotiation}}, a key step is injecting a \texttt{FOD}
packet with a low reference Q-factor. This strategy exploits the charger's FOD
check mechanism, which compares the measured Q-factor against the reference
value provided by the RX device. By setting a low threshold, the charger is
misled into passing the FOD check and issuing an \texttt{ACK} response.
Subsequently, we request further details from the charger, such as its ID and
charging capabilities, by injecting general request (\texttt{GRQ}) and specific request (\texttt{SRQ}) packets.
After \texttt{\textbf{negotiation}}, the charger is successfully directed to the power
transfer stage with the extended protocol at $t_1$. At this point, the power
transfer rate is adjusted and kept high through the injection
of tailored \texttt{CE} and \texttt{RP} packets, heating up and potentially
damaging foreign objects.

\subsubsection{Attack Evaluations}
\begin{figure}[htbp]
    \centering
    \includegraphics[width=.325\columnwidth]{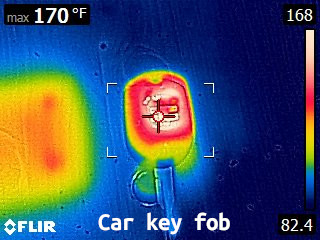}
    \includegraphics[width=.325\columnwidth]{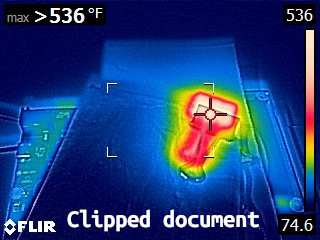}
    \includegraphics[width=.325\columnwidth]{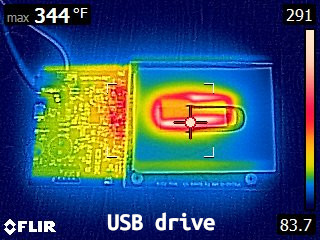}\\
    \includegraphics[width=.325\columnwidth]{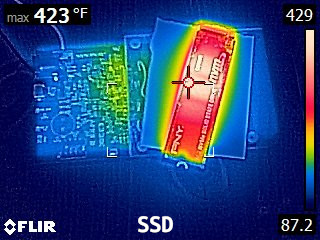}
    \includegraphics[width=.325\columnwidth]{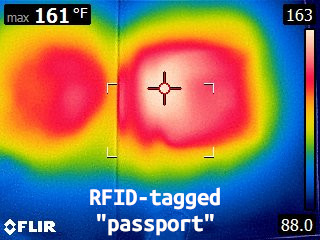}
    \includegraphics[width=.325\columnwidth]{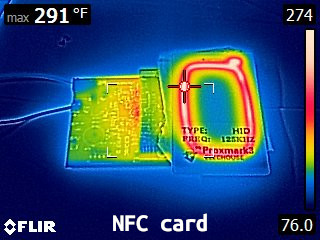}\\
    \includegraphics[width=.242\columnwidth]{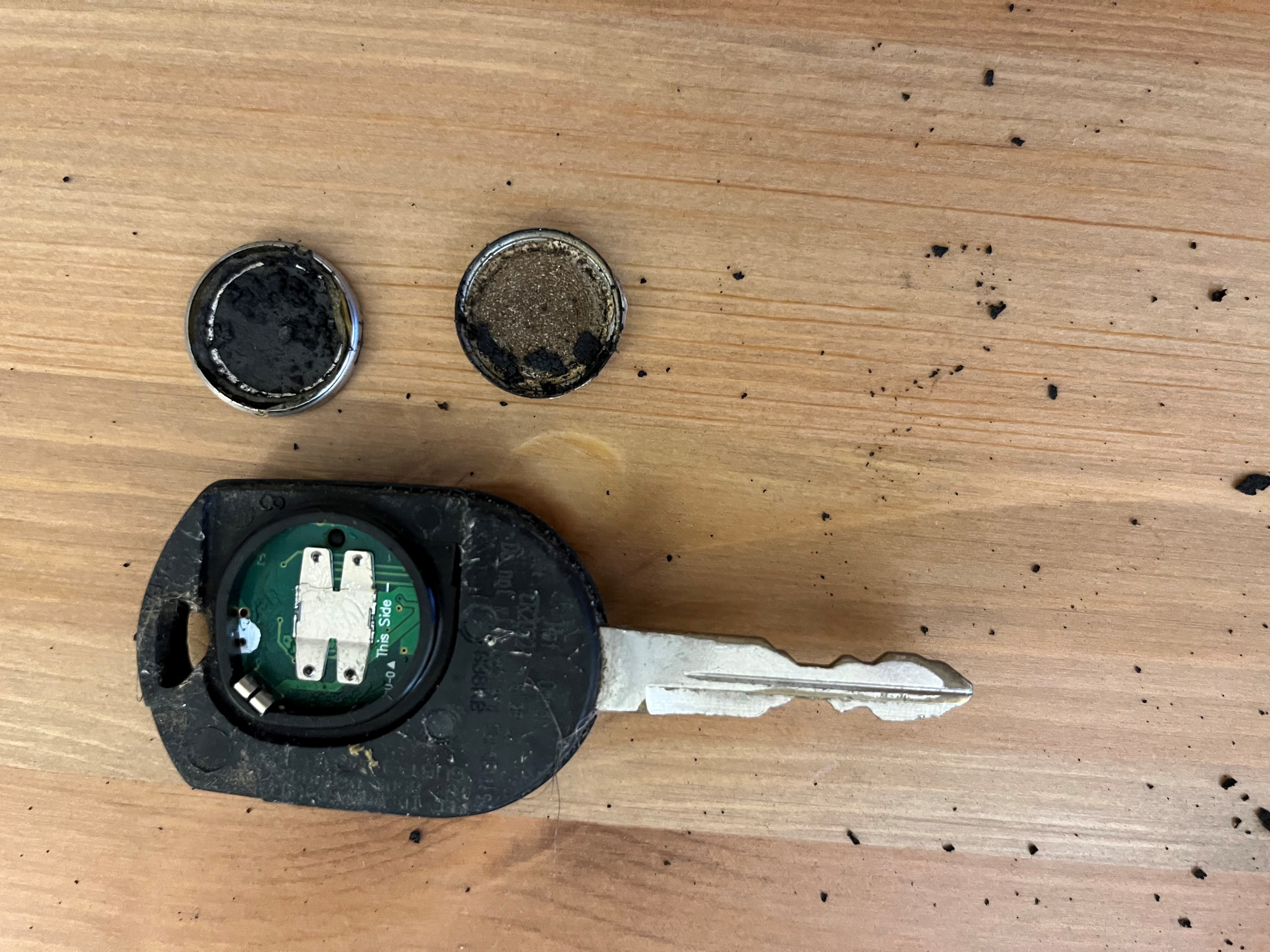}
    \includegraphics[width=.242\columnwidth]{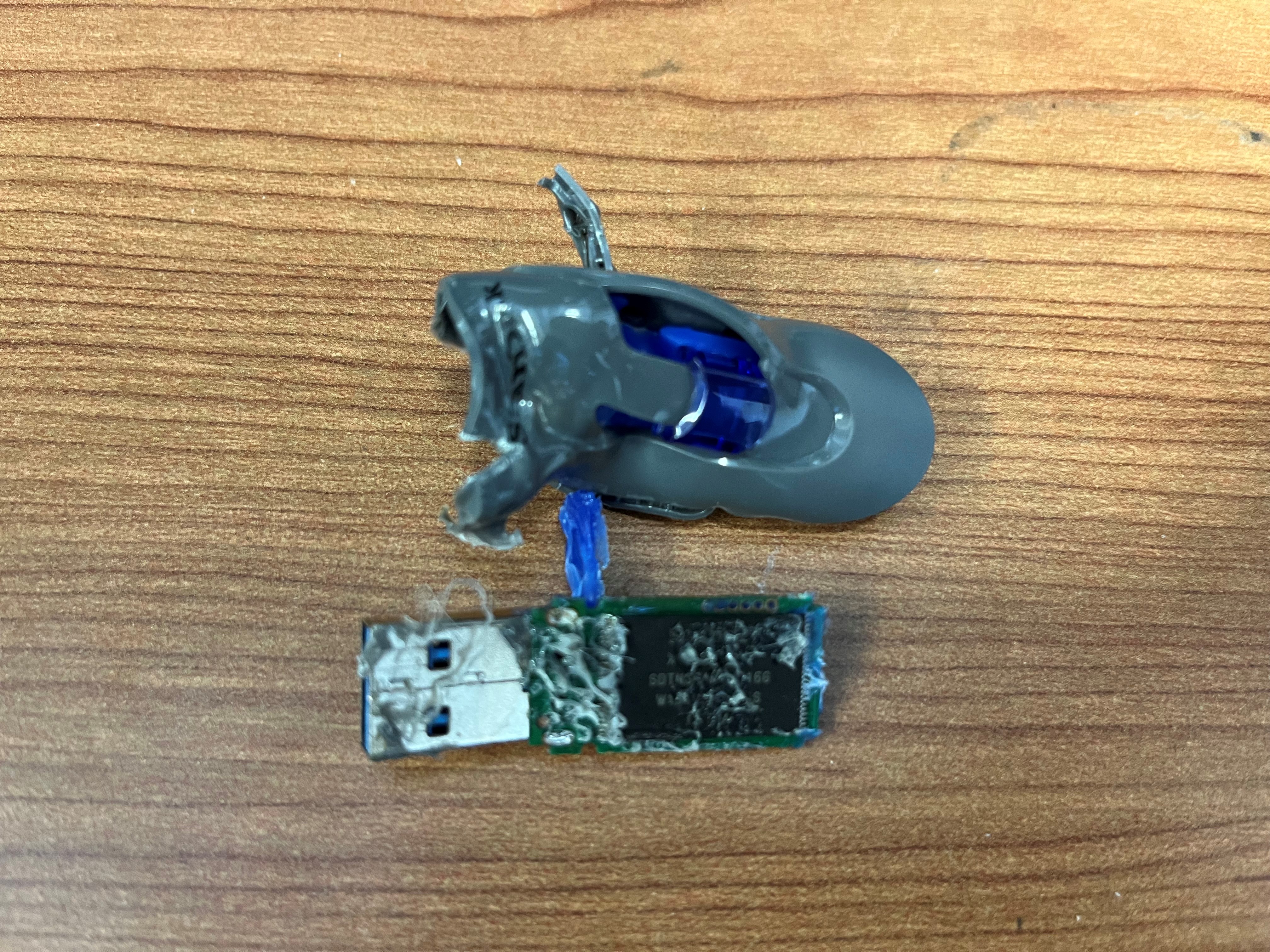}
    \includegraphics[width=.242\columnwidth]{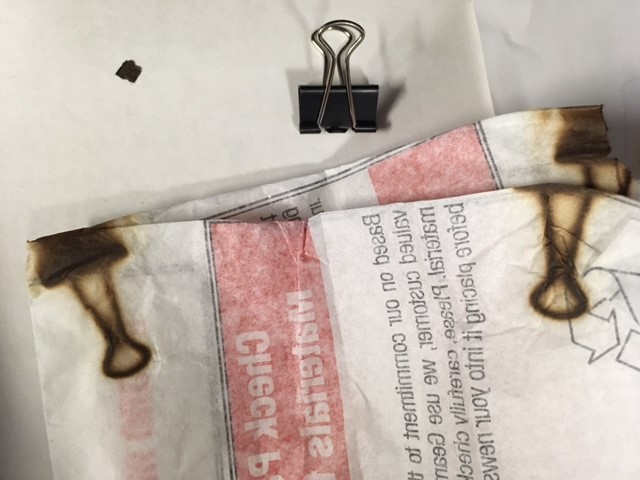}
    \includegraphics[width=.242\columnwidth]{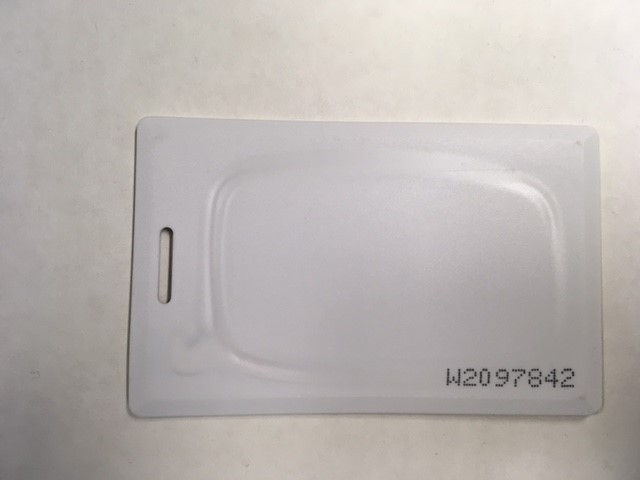}\\
    \caption{Thermal images and visible damages on different targets}
    \label{fig:fod_heatmap}
\end{figure}

We carried out the attacks on six common personal items, initiating power transfer and maintaining the maximum charging power until visible damage occurs or the maximum temperature is sustained for two hours. Our evaluations, as shown in Figure~\ref{fig:fod_heatmap}, reveal some concerning outcomes:\\
\textbf{Key Fob:} Upon initiating power transfer to a car key fob placed on the
charging pad, the battery inside reached a critical temperature. As a result,
the key fob didn't merely overheat. Instead, it detonated and caused the
disintegration of the device in an explosive display. \\
\textbf{Paper Clips:} The temperature exceeded 536°F when heated, which can
potentially damage or destroy important documents affixed by these clips.\\
\textbf{USB Drive:} The high temperature caused significant damage to the USB
drive and the memory chip, making the contained data unrecoverable.\\
\textbf{Solid-State Drive (SSD):} SSD is commonly found on laptops and can be
accidentally placed on the charging pad. We find that our attack can overheat
the controller and flash of SSD into unrecoverable states thus rendering
it to suffer data loss.\footnote{The SSD is expected to be more susceptible to high temperature when actively operating in a laptop because the maximum operating temperature specified for SSD is 149 $^\circ$F.}\\
\textbf{Passport and NFC Cards:} Personal identification documents often contain RFID tags as identification chips. Similarly, NFC cards
are often used as security tokens for verification. However, when these items are
accidentally left on the charging pad, the strong magnetic field generated by
the charger can immediately destroy these identification tokens.

We tested each charger for its destructive potential on the objects and
measured the maximum charging power achievable when transferring power
to a paper clip. The results listed in Table~\ref{tab:evaluation_fod} show that all chargers can readily destroy RFID tags and NFC cards.
The damage potential increases with the increased charging power. Even if some
chargers do not directly damage certain objects, they can generate temperatures
exceeding the safe limits for components like SSDs and USB drives, thereby
causing permanent data loss.
\begin{table}[htbp]
    \centering
    \small
    \caption{Foreign object destruction ability}
    {\fontfamily{ppl}\selectfont
        \begin{tabular}{|c|c|c|c|c|c|c|}
            \hline
            ID.              & SSD    & USB    & KFB    & NFC    & RFID   & PWR (W) \\
            \hline
            \rowcolor{5W} 1  & \xmark & \xmark & \xmark & \cmark & \cmark & 6       \\
            \rowcolor{10W} 2 & \xmark & \xmark & \xmark & \cmark & \cmark & 5       \\
            \rowcolor{10W} 3 & \xmark & \xmark & \xmark & \cmark & \cmark & 7.9     \\
            \rowcolor{10W} 4 & \xmark & \xmark & \cmark & \cmark & \cmark & 9.3     \\
            \rowcolor{10W} 5 & \xmark & \xmark & \xmark & \cmark & \cmark & 5.5     \\
            \rowcolor{10W} 6 & \xmark & \xmark & \cmark & \cmark & \cmark & 9.2     \\
            \rowcolor{15W} 7 & \cmark & \cmark & \cmark & \cmark & \cmark & 19      \\
            \rowcolor{15W} 8 & \cmark & \cmark & \cmark & \cmark & \cmark & 18      \\
            \rowcolor{15W} 9 & \cmark & \cmark & \cmark & \cmark & \cmark & 15      \\
            \hline
        \end{tabular}
    }
    \label{tab:evaluation_fod}
    \vspace*{-.15in}
\end{table}

%% file: files/discussion.tex
\section{Discussion}\label{sec:discussion}

In this section, we discuss the practicality and stealthiness of our attacks,
compare our work with state-of-the-art research, and provide insights for diverse charging
protocols. We also propose countermeasures to mitigate the risks of our attacks.

\subsection{Comparison With Prior Works}

To clarify the uniqueness of \VoltSchemer, we conducted a detailed comparison
with state-of-the-art wireless charger manipulation attacks~\cite{Wu:2021:ACSAC,
Dai:2023:SP}. This comparative analysis, outlined in
Table~\ref{tab:comp_with_prior}, focuses on the practical implementation
aspects and the specific attack capabilities of these methods. In-depth
discussions of these two aspects are provided in the remaining part of this
section.

\definecolor{bad}{HTML}{AE2012}
\begin{table}[htbp]
    \small
    \centering
    \caption{Comparison with state-of-the-art works}
    \label{tab:comp_with_prior}
    {\fontfamily{ppl}\selectfont
        \begin{tabular}{|c|c|c|c|c|c|c|c|}
            \hline
            \multirow{2}{*}{Work}            & \multicolumn{4}{c|}{Practicality} & \multicolumn{3}{c|}{Attacks}                                                                                                                         \\
            \cline{2-8}
            \rule{0pt}{2.5ex}                & \faWrench                         & \faExchange                  & \faEyeSlash           & \faCartArrowDown      & \faMicrophone         & \faBattery[4]         & \faChainBroken        \\
            \hline
            Qi Hijacking\cite{Wu:2021:ACSAC} & \cmark                            & \cmark                       & \textcolor{bad}\xmark & \textcolor{bad}\xmark & \textcolor{bad}\xmark & \cmark                & \textcolor{bad}\xmark \\
            \hline
            Wormheart\cite{Dai:2023:SP}      & \textcolor{bad}\xmark             & \textcolor{bad}\xmark        & \cmark                & \textcolor{bad}\xmark & \cmark                & \textcolor{bad}\xmark & \textcolor{bad}\xmark \\
            \hline
            Parasite \cite{Dai:2023:SP}      & \cmark                            & \cmark                       & \textcolor{bad}\xmark & \textcolor{bad}\xmark & \cmark                & \textcolor{bad}\xmark & \textcolor{bad}\xmark \\
            \hline
            \VoltSchemer                     & \cmark                            & \cmark                       & \cmark                & \cmark                & \cmark                & \cmark                & \cmark                \\
            \hline
        \end{tabular}
    }

    \faWrench: Feasible installation, \faExchange: Versatility, \faEyeSlash:
    Stealthy modification, \faCartArrowDown: COTS evaluations, \faMicrophone:
    Voice assistant manipulation, \faBattery[4]: Charging manipulation,
    \faChainBroken: Foreign object destruction

\end{table}

\begin{figure}[htbp]
    \centering
    \includegraphics[width=\columnwidth]{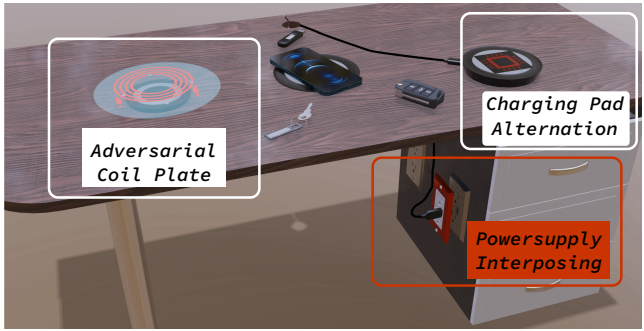}
    \caption{Three wireless charger manipulation methods}
    \label{fig:comparion_implementation}
\end{figure}

\noindent\textbf{Comparison of Implementation Practicality}
Figure~\ref{fig:comparion_implementation} shows three different methods of wireless charger manipulation attacks: 
\Circled{\textbf{1}} adversarial coil plate insertion, 
\Circled{\textbf{2}} charging pad alternation, 
and \Circled{\textbf{3}} power supply interposing.

The ``Wormheart'' attack~\cite{Dai:2023:SP} involves installing customized
firmware in the charger, usually by modifying or replacing its MCU. However, as
detailed in Appendix~\ref{appx:chip_size}, the MCU's small size and dense
integration on the charger board make malware installation infeasible. Moreover,
this method's versatility is limited as each distinct charging system
necessitates a uniquely customized malware. The work by Wu \etal
\cite{Wu:2021:ACSAC} and the ``Parasite'' voice assistant manipulation
attack~\cite{Dai:2023:SP} both require inserting adversarial coils over the
genuine wireless charger. Because users must place devices on the adversarial
coil for each charging session, such frequent interaction increases the chance
of discerning the anomalies, thereby undermining the attack's stealthiness. Our
\VoltSchemer\ attacks employ IEMI on the power supply to control the charger,
requiring only an intermediary device connection to the power adapter. While
both \VoltSchemer\ and adversarial coil methods involve adding a device, ours is
more covert. Primarily, our method capitalizes on the infrequent inspection of
power adapters and charging cables, in line with wireless charging's core
principle of minimal wire interaction. Furthermore, replicating a standard power
port is more viable, owing to the common, simple design of regular outlets. In
addition to these advantages, our approach's versatility is demonstrated by
testing on 9 different wireless chargers, including COTS devices, a significant
expansion from previous works~\cite{Wu:2021:ACSAC,Dai:2023:SP} that only
assesses a single evaluation board charger.

\noindent\textbf{Comparison of Attack Capability} Our research outweighs
state-of-the-art works in both the breadth and depth of evaluations concerning
three attack capabilities. The voice assistant manipulation attack in
\cite{Dai:2023:SP} is narrowly focused on a single custom-built wireless
charger, only testing the activation of voice assistants. Our \VoltSchemer\
approach broadens this scope significantly by evaluating 9 varied COTS wireless
chargers with 6 different common voice commands. This not only proves the
versatility of \VoltSchemer\ across various hardware configurations but also
uncovers deeper insights into the security risks associated with voice assistant
manipulation attacks, highlighting the importance of comprehensive security
measures in wireless charging technologies. Wu et al.'s work
\cite{Wu:2021:ACSAC} demonstrates the impact of injected \texttt{CE} packets on
charging power, but didn't progress to practical attacks. Our \VoltSchemer\
evaluations reveal that altering \texttt{CE} packets alone is ineffective
against modern smartphones' overcharging protections. Leveraging an in-depth
understanding of Qi wireless charging protocols, we develop a practical power
toasting attack with more skillfully controlled implementations. Our tests
confirm that \VoltSchemer\ can circumvent three protective measures, causing
dangerously high temperatures in smartphones, thereby demonstrating a deeper
insight into the attack's causes and impacts. Moreover, we introduce an unprecedented attack
scenario in existing research. Our extensive evaluations show that \VoltSchemer\
can manipulate wireless chargers to breach the protections of Qi standard, causing
damage to metallic foreign objects, showcasing the potential for significant
property loss and safety hazards.

\subsection{Insights for Diverse Charging Protocols}

The core issue facilitating our attacks is the insufficient noise suppression in certain frequency bands, leaving systems vulnerable to interference even if they meet existing EMC/EMI standards. This gap makes all wireless charging technologies potentially vulnerable to interference-based attacks, particularly high-power systems like electric vehicle (EV) wireless charging. Despite the nascent stage of EV wireless charging standards and efforts to incorporate safety measures, our research demonstrates the significant risks of system compromise, including property damage and threats to human safety. Our findings reveal the urgent need for improved protective measures against such IEMI interference, pointing to the critical importance of safeguarding wireless charging infrastructure from these sophisticated threats.

\subsection{Countermeasures}

\begin{figure}[htbp]
    \centering
    \includegraphics[width=\columnwidth]{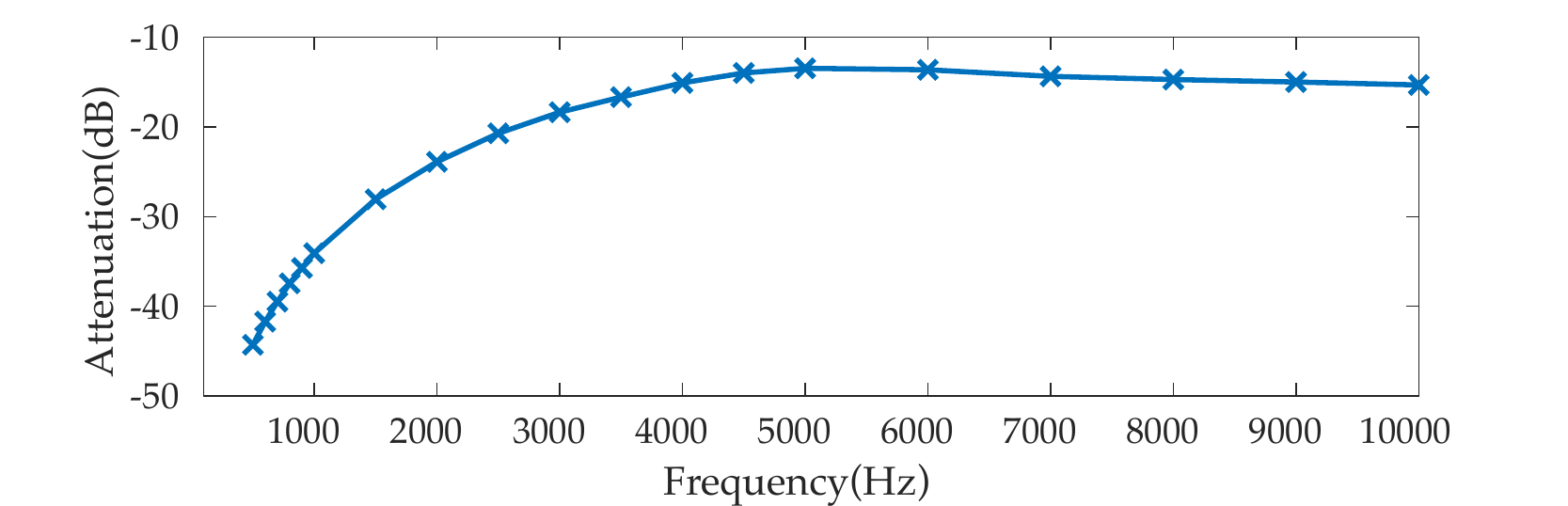}
    \caption{A DC/DC converter's noise attenuation for input voltage as a function of frequency}
    \label{fig:dcdcatten}
\end{figure}

A practical countermeasure to our attacks involves integrating noise suppression
components, such as additional DC/DC converters, to remove noise in the input
voltage. To validate this approach, we connect a DC/DC converter to the input power
port of a Renesas P9242 wireless charger and assess the attenuation of injected
noise. By injecting voltage noises across frequencies ranging from 500 Hz to
10kHz and measuring the voltages both before and after the DC/DC converter, we
quantify the attenuation level. As Figure~\ref{fig:dcdcatten} illustrates, the
converter achieves a minimum noise reduction of 15 dB, with more substantial
attenuation at lower frequencies. This additional converter effectively
mitigates all three attacks. However, this solution comes with trade-offs. For
instance, it increases the charger's cost, size, weight, and failure
rate. Moreover, the additional components also increase the power consumption
and pose more thermal stress challenges.

An alternative countermeasure involves real-time monitoring the voltage waveform DC bus. 
If the charger detects abnormal noises, which may indicate IEMI injection, it can respond by triggering alarms or shutting down to avoid further damage. 
However, the cost implications of implementing this mitigation may also pose a challenge for low-cost devices.

%% file: files/related_work.tex
\section{Related Work}

\subsection{Attacks during Charging}
Smart devices often exchange information with chargers during the charging processes via
USB cables, which also help to transfer files or install applications. The charging process
can be exploited for eavesdropping, as changes in power consumption can be
detected through the charging channel. 

With \textit{\textbf{Wired Charging}}, studies have shown that malicious
charging cables can be used to control mobile devices and install malicious
applications~\cite{lau2013mactans, nohl2014badusb, omgteam}. Certain techniques
can bypass the port lock mechanism, inject voice commands~\cite{Wang:2022:NDSS},
or inject touch events onto touchscreens~\cite{Jiang:2022:SP}. There are also
techniques to procure sensitive information from the charged devices, like screenlock
passwords~\cite{shiroma2017threat, cronin2021charger}, browsing
activities~\cite{yang2016inferring}, and installed
applications~\cite{chen2017powerful}. \textit{\textbf{Wireless charging}}, while
popular due to its cordless design, presents new challenges. It has been
demonstrated that wireless charging can also be vulnerable to side channel
attacks~\cite{La:2021:CCS, ni2022uncovering}. Vulnerabilities in the Qi wireless
charging protocol have been exposed, which can be exploited to inject malicious
charging commands and eavesdrop using an externally placed
coil~\cite{wu2020security, Wu:2021:ACSAC}. Further improvements in eavesdropping
attacks have been made by measuring the power consumption of the wireless
charger~\cite{liu2022privacy}. There are also techniques that use a customized
wireless charging coil to induce magnetic interference and inject voice
commands~\cite{Dai:2023:SP, dai2022inducing}.

\subsection{Inaudible Voice Injection Attacks}
There are many well-known attacks on microphones to manipulate the sensed voice
on smart devices and inject malicious voice commands. Among these voice
injection attacks, two main categories of attacks are often discussed.

\textit{\textbf{Indistinguishable Voice Injection}} generates malicious audio
that can be interpreted by speech recognition systems but not by humans. This
attack is demonstrated by Vaidya \etal\cite{vaidya2015cocaine} and Carlini
\etal\cite{carlini2016hidden}, further improved by Yuan
\etal\cite{yuan2018commandersong} by embedding voice commands into songs.
Sch"{o}nherr \etal\cite{Schoenherr2019} and Abdullah
\etal\cite{abdullah2019practical} further refined the attack for broader use and
practicality. Although researchers use several means to generate better
malicious audio, this type of attack still relies on the fact that an audible
voice carrier is needed, which is a hard requirement.

\textit{\textbf{Inaudible Voice Injection}} produces voice signals only
detectable by microphones. Wang \etal\cite{zhang2017dolphinattack},
Sugawara\cite{Sugawara:2020:USENIX}, and Roy \etal\cite{roy2018inaudible}
proposed using ultrasonic frequency carrier signals, laser signals, and
ultrasound speaker arrays for such attacks. Ji \etal\cite{ji2021capspeaker} used
an implanted capacitor for this purpose. Dai \etal\cite{Dai:2023:SP,
dai2022inducing} and Wang \etal\cite{Wang:2022:NDSS} demonstrated this attack
can be executed via a wireless charger or a charging cable.

%% file: files/ethical_considerations.tex
\section{Ethical Considerations}
\textbf{Responsible Disclosure} We have contacted vendors to report the
identified vulnerabilities, including NXP, Renesas, Infineon, ST, Wireless Power Consortium, etc.
Countermeasures that can be employed by hardware vendors are under discussion
and will be further disclosed in the future.

\noindent
\textbf{IRB Approval} 
The University of Florida Institutional Review Boards have approved this research.
The IRB approval number is ET00020284.

\noindent 
\textbf{Impact on Power Grid Integrity} 
Following reviewers' recommendations,
we evaluated our experiment's potential impact on the power grid's integrity. We can ascertain that the impact is negligible.
This is largely due to the power adapter's noise-isolation design and the low-power interference signals used. 
However, future research involving IEMI should proactively and thoroughly assess the potential impact on the integrity of power grid, particularly in scenarios where interference is injected closer to the grid or with higher intensity.

\noindent
\textbf{Safety Measures} In our study, certain experiments posed risks of
battery fires and explosions. To address these concerns, we set up a controlled
environment to ensure safety. The experiments took place in a clean,
non-flammable area, equipped with adequate ventilation to prevent the
accumulation of hazardous gases. Protective barriers were installed around the
Device Under Test (DUT) to contain any fragments from potential explosions.
Moreover, we ensured the availability and accessibility of specialized fire
extinguishers, specifically designed for handling electrical and chemical fires,
as a crucial safety measure.

%% file: files/conclusion.tex
\section{Conclusion}
In this paper, we identified vulnerabilities of wireless chargers that enable
the implementation of \VoltSchemer, a set of powerful and practical active
attacks against COTS wireless chargers. Exploiting voltage interference on the
power adapters' output voltage, \VoltSchemer\ can manipulate the chargers to
perform malicious activities like injecting inaudible voice commands to control
voice assistants, overheating the charged devices, and destroying metallic
foreign objects. Comprehensive evaluations of top-selling wireless chargers
confirm the effectiveness and practicality of \VoltSchemer\ attacks.

%% file: files/acknowledgement.tex
\section*{Acknowledgement}
We appreciate the reviewers and the shepherd for their insightful comments and suggestions. 
This work was supported partially by the National Science Foundation under award numbers 1818500 and 1916175, and partially by the gift donation from Intel.

%% file: files/appendix-burned_marks.tex
\section{Attacking Practicality Discussion}\label{appx:chip_size}
\begin{figure}[htbp]
    \centering
    \includegraphics[width=.5\columnwidth]{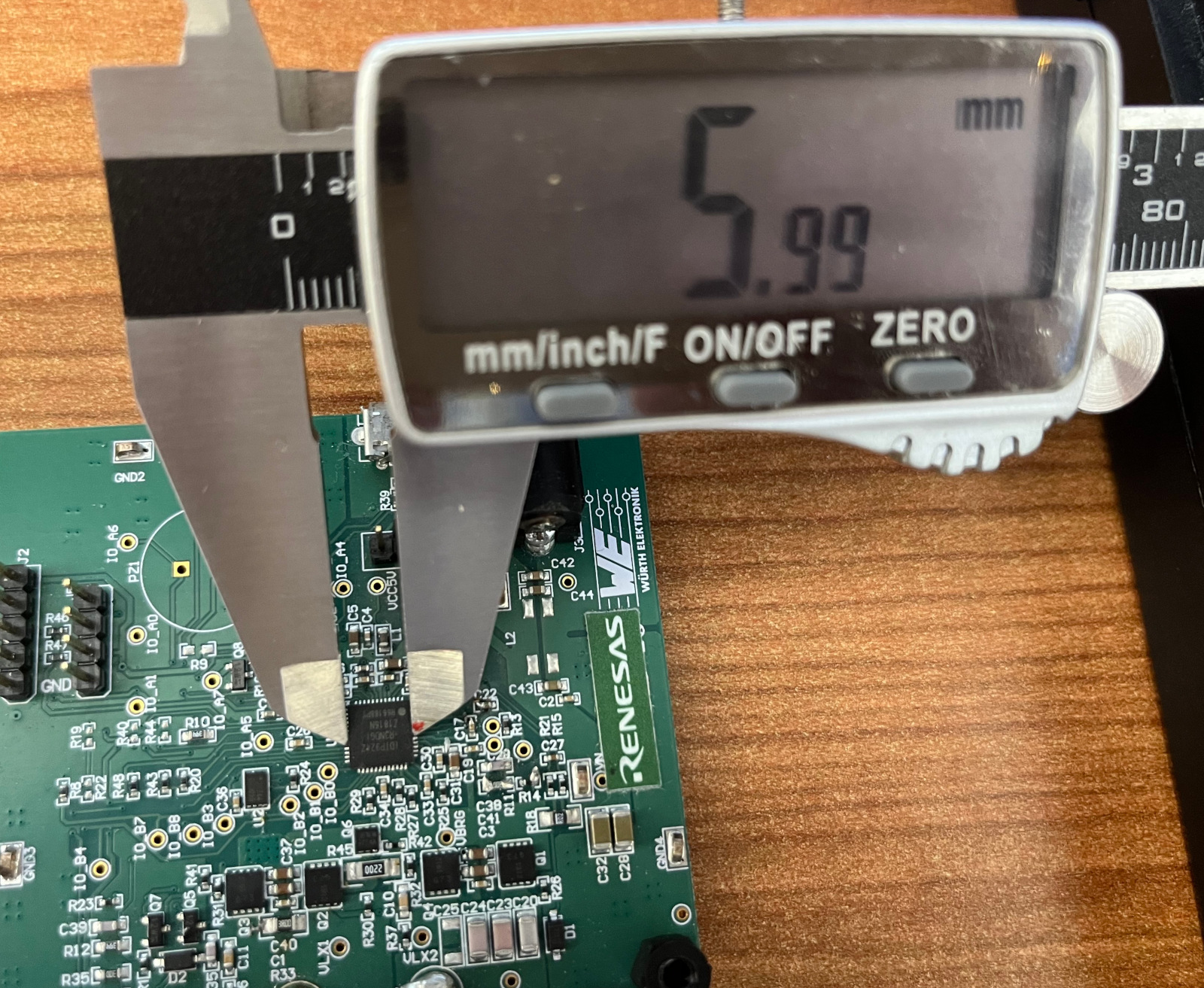}
    \caption{Microcontroller chip on the wireless charger}
    \label{fig:controller_chipsize}
\end{figure}

Figure~\ref{fig:controller_chipsize} shows a microcontroller chip in a wireless charger.
Due to its compact size and high level of integration on the board, malicious charging pad modifications requiring chip replacement are difficult to perform.
This feature limits the practicality of the ``Wormheart'' attack.

Despite their small size, such chips are capable of performing complex computations, including processing voltage traces, decoding Qi messages, and generating control signals for power signal modulation.
Thus, if mass production is needed, the size of our prototype \VoltSchemer\ can be significantly reduced by substituting the AD2 with a chip at this scale.

\section{Inverter Output Voltage}\label{appx:inverter_signal}

\begin{figure}[htbp]
    \centering
    \includegraphics[width=.8\columnwidth]{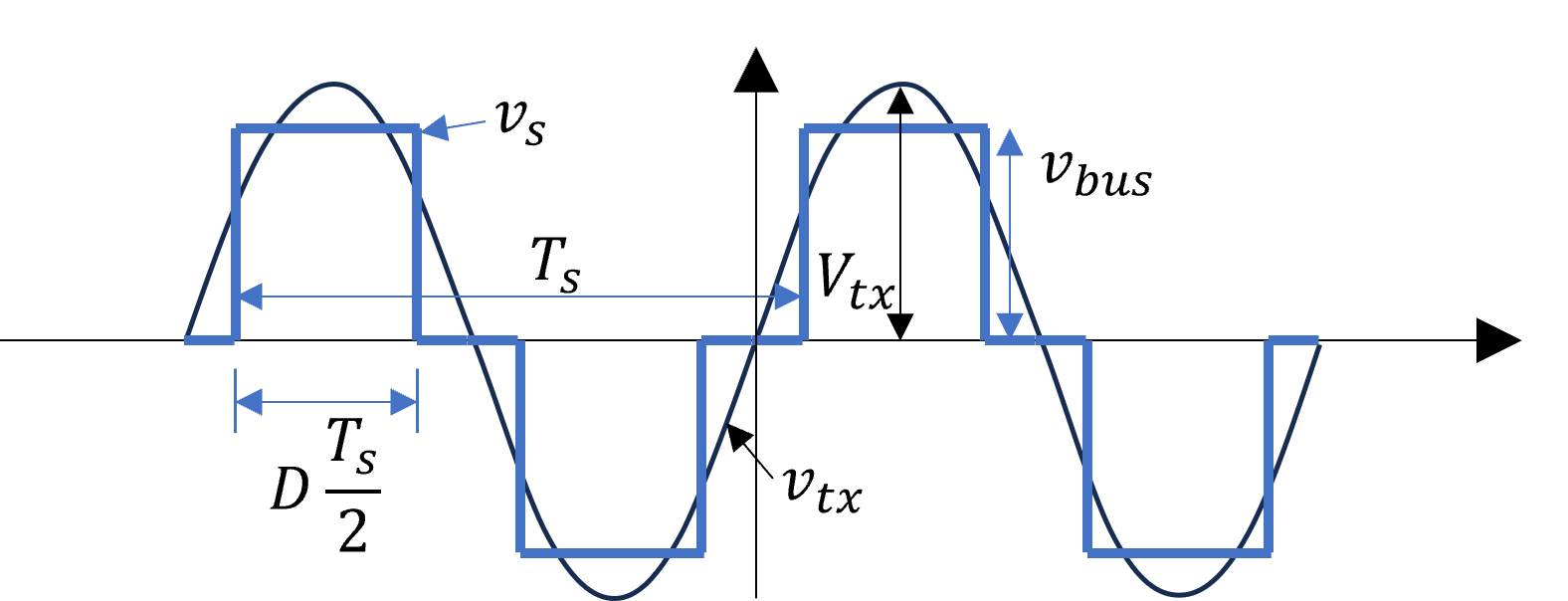}
    \caption{Waveform of inverter output voltage}
    \label{fig:inverter_waveform}
\end{figure}

When an inverter operates at a switching frequency $f_p$ with a duty cycle $D$, the waveform $v_s(t)$ of the inverter output voltage is illustrated in Figure~\ref{fig:inverter_waveform}. $v_s(t)$ is mathematically described by Equation~\ref{eq:wave_switching} over the interval $\left[-\frac{T_s}{2}, \frac{T_s}{2}\right]$, where $T_s$, the period of the switching pattern, is defined as $\frac{1}{f_p}$.

\begin{equation}\label{eq:wave_switching}
v_s(t) = 
\begin{cases}
v_{\text{bus}} & \frac{T_s}{4}(1 - D) < t < \frac{T_s}{4}(1 + D) \\
-v_{\text{bus}} & -\frac{T_s}{4}(1 + D) < t < -\frac{T_s}{4}(1 - D) \\
0 & \text{otherwise}
\end{cases}
\end{equation}

If the voltage $v_{tx}$ corresponds to the fundamental harmonic of $v_s(t)$ at frequency $f_p$ and the amplitude of the fundamental of $v_s(t)$ at $f_p$ is $V_{tx}$, when filtered through a resonance tank that only retains the fundamental component, the voltage $v_{tx}$ can be expressed as:
\begin{equation}
\begin{split}
    v_{tx}(t) &= V_{tx} sin(2\pi f_pt)\\
    &= \frac{4}{\pi}\sin{(\frac{\pi}{2}D)} v_{bus} \sin(2\pi f_p t)
\end{split}
\end{equation}